\documentclass[10pt,a4paper,review]{elsarticle}
\usepackage[left=1.0in, right=1.0in, top=1.0in]{geometry}
\usepackage[table]{xcolor}
\usepackage{enumitem}
\usepackage{tabularx}
\usepackage{ragged2e}

\usepackage{tikz}
\usetikzlibrary{arrows.meta}

\tikzset{brace/.style={decorate, decoration={brace}},
  brace mirrored/.style={decorate, decoration={brace,mirror}},
}

\usepackage[utf8]{inputenc} 
\usepackage[T1]{fontenc}    
\usepackage{hyperref}       
\usepackage{url}            
\usepackage{booktabs}       
\usepackage{amsfonts}       
\usepackage{nicefrac}       
\usepackage{microtype}      

\usepackage{amssymb}
\usepackage{amsmath}
\usepackage{lineno,hyperref}

\usepackage{bm}
\usepackage{subcaption}
\usepackage{graphicx}
\usepackage{svg}
\usepackage{rotating}
\usepackage{lscape}
\usepackage[title]{appendix}
\usepackage{float}

\usepackage{amsthm}
\usepackage{listings}
\usepackage{xcolor}
\usepackage{siunitx}

\usepackage{pifont}
\usepackage{lineno}

\usepackage{import}
\usepackage{physics}
\usepackage{comment}
\usetikzlibrary{positioning}

    \DeclareFontFamily{OT1}{pzc}{}
\DeclareFontShape{OT1}{pzc}{m}{it}{<-> s * [1.10] pzcmi7t}{}
\DeclareMathAlphabet{\mathpzc}{OT1}{pzc}{m}{it}
\biboptions{sort&compress}

\usepackage{algpseudocode,algorithm,algorithmicx}

\makeatletter
\def\ALG@special@indent{%
    \ifdim\ALG@thistlm=0pt\relax
        \hskip-\leftmargin
    \else
        \hskip\ALG@thistlm
    \fi
}
\newcommand{\Begin}[1]{\item[]\noindent\ALG@special@indent \textbf{Begin:}\ #1}
\newcommand{\Establish}[1]{\item[]\noindent\ALG@special@indent \textbf{Establish:}\ #1}
\newcommand{\End}{\item[]\noindent\ALG@special@indent \textbf{End}}
\makeatother
\algblock[Name]{Start}{End}
\algblockdefx[NAME]{START}{END}%
    [2][Unknown]{Start #1(#2)}%
    {Ending}

\usepackage{nccmath}

\usetikzlibrary{arrows}

\newcommand{\rev}[1]{\textcolor{black}{#1}}
\makeatletter
\newcommand*{\compress}{\@minipagetrue}
\makeatother

\renewcommand\citep[1]{%
    \citeauthor{#1}~(\citeyear{#1})%
}

\begin{document}

\title{NN-EVP: A physics informed neural network-based elasto-viscoplastic framework for predictions of grain size-aware flow response under large deformations }
\author[CORNELL]{Adnan Eghtesad\corref{cor1}}
 \ead{ae386@cornell.edu}
\author[CORNELL]{Jan Niklas Fuhg}
\author[CORNELL,CORNELL2]{Nikolaos Bouklas}
\cortext[cor1]{Corresponding author.}

\address[CORNELL]{Sibley School of Mechanical and Aerospace Engineering, Cornell University,  NY 14853, USA}
\address[CORNELL2]{  Center for Applied Mathematics,
  Cornell University,
   NY 14853, USA  }

\begin{abstract}
We propose a physics informed, neural network-based elasto-viscoplasticity (NN-EVP) constitutive modeling framework for predicting the flow response in metals as a function of underlying grain size. The developed NN-EVP algorithm is based on input convex neural networks as a means to strictly enforce thermodynamic consistency, while allowing high expressivity towards model discovery from limited data. It utilizes state-of-the-art machine learning tools within PyTorch's high-performance library providing a flexible tool for data-driven, automated constitutive modeling.
To test the performance of the framework, we generate synthetic stress-strain curves using a power law-based model with phenomenological hardening at small strains and test the trained model for strain amplitudes beyond the training data. Next, experimentally measured flow responses obtained from uniaxial deformations are used to train the framework under large plastic deformations. Ultimately, the Hall-Petch relationship corresponding to grain size strengthening is discovered by training flow response as a function of grain size, also leading to efficient extrapolation.  The present work demonstrates a successful integration of neural networks into elasto-viscoplastic constitutive laws, providing a robust automated framework for constitutive model discovery that can efficiently generalize, while also providing insights into predictions of flow response and grain size-property relationships in metals and metallic alloys under large plastic deformations.  
\end{abstract}


\maketitle

\break
\section{Introduction}
 
Understanding the flow response and viscoplastic behavior of metals incorporating strain rate sensitivity effects, is essential for the predictions of the mechanical behavior, improving the material performance, and designing novel alloys, critical towards revolutionizing the objectives of the aerospace and automotive manufacturing industries. Owing to their polycrystalline nature, the anisotropic flow response of metals under large deformations heavily depends on the average grain size that forms the underlying microstructure. Grain size strengthening, referred to as the Hall-Petch effect \cite{HANSEN2004801,dunstan2014grain,figueiredo2023seventy}, is associated with the grain boundary resistance to the crystallographic deformation mechanisms and dislocation motions \cite{eghtesad2022density,figueiredo2021deformation}. Authors in \cite{agius2022crystal} implemented a crystal plasticity model that accounts for grain size effects and slip system interactions on the deformation of austenitic stainless steels. The crystal plasticity finite element method (CPFEM) is used to study the grain size and morphology effects on yield strength \cite{lakshmanan2022crystal}. In addition to crystal plasticity modeling, atomistic simulations are well utilized to study the Hall-Petch relationships in advanced alloys \cite{zhang2020inverse,zhang2020quantifying}.   

High-performance computing (HPC) has improved the efficiency of numerical techniques over the last few decades. However, despite acceleration of HPC simulations, predicting multi-scale material deformations is still a time-consuming task and limited by parallel scalability and hardware performance. 
To address this issue, recent research has focused on the integration of genetic algorithms (GA), machine learning (ML), and deep learning (DL) to facilitate automated constitutive modeling which has the potential to expedite multi-scale simulations \cite{le2015computational,fuhg2021model,kalina2023fe}. In particular, they have been used to model hyperelasticity \cite{vlassis2020geometric,fuhg2022physics,klein2022polyconvex,kalina2022automated,fuhg2022learning,tacc2023benchmarking}, viscoelasticity \cite{tacc2023data,as2023mechanics,flaschel2023automated,upadhyay2023physics} and multiphysics problems \cite{kadeethum2021framework,klein2022finite}.

Lately, neural networks (NN) and convolutional neural networks (CNN) have also been utilized for a variety of applications in modeling viscoplastic deformations in macro and micro scales \cite{yao2021hybrid}. A recurrent neural network (RNN) model was proposed as a computationally-efficient surrogate for crystal plasticity simulations \cite{bonatti2022cp}. A new NN-based crystal plasticity algorithm was presented for FCC materials and its application to non-monotonic strain paths \cite{ibragimova2021new} followed by a CNN-based CPFEM  model to predict the localized viscoplastic deformation in aluminum alloys \cite{ibragimova2022convolutional}. Some recent work has implemented NN and CNN to predict yield surfaces from microstructural images and crystal plasticity simulations \cite{heidenreich2023modeling,nascimento2023machine}. A machine learning-enabled crystal plasticity model with dislocation density hardening was developed to identify stress and strain localizations under large viscoplastic deformations \cite{eghtesad2023machine}. An input convex NN (ICNN) framework was presented for the prediction of texture-dependent macroscopic yield functions from crystal plasticity simulations \cite{fuhg2022machine}. On the continuum level, NN algorithms have been implemented in finite element (FE) solvers to replace classical history-dependent constitutive models to obtain nonlinear structural responses \cite{stoffel2019neural,huang2020machine,stoffel2020deep,tandale2022physics,benabou2021implementation}.     
 
While the literature addresses a wide range of machine and deep learning applications related to elastoplasticity and viscoplasticity in the context of big data, only a few rely on the more realistic case of limited-data availability of macroscopic observations from measured tensile tests. Furthermore, studies that focus on viscoplasticity in the low- and limited-data regimes, only utilize neural networks for parameter estimation of established phenomenological constitutive models rather than model discovery. The latter can focus on establishing automated frameworks that remove the need for a particular phenomenological parametrization. To address this shortcoming, and allow robust performance in low- and limited-data regimes, physics-based ML algorithms have been proposed aiming to directly enforce physical laws, thermodynamic principles and also established domain knowledge \cite{brodnik2023perspective}.  
Motivated by the earlier work of \cite{vlassis2021sobolev}, recent work in \cite{fuhg2023modular}, enables modular ML-based elastoplasticity in the context of limited data using thermodynamics-aware and mechanistically informed neural networks  offering a hybrid framework with each component of the model being selective as classical phenomenological or a data-driven model depending on the data availability, allowing for trustworthy prediction and generalization. 

The present work builds upon the work in \cite{fuhg2023modular} and proposes a novel NN-based framework for modeling the elasto-viscoplastic (NN-EVP) response of metals as a function of grain size where strain rate sensitivity effects are taken into consideration. The proposed algorithm and underlying constitutive model consider the coupled elasto-viscoplastic response in contrast to the studies where elastic and plastic regimes are treated separately. In addition, for consistency with the thermodynamic laws and mechanistic assumptions, neural network architectures are designed to enforce monotonicity and convexity requirements. 
The framework allows for the discovery of laws describing the rate-sensitive viscoplastic flow response of metals and alloys, enabling a predictive infrastructure for the flow response as a function of strain amplitude and grain size. The present paper is structured as follows. Section 2 starts with a brief review of the thermodynamics-based modular formulation, based on dual potentials, as well as particular functional forms, resulting from mechanistic assumptions, that describe the rate-sensitive viscoplastic flow response in metals.  Afterward, the key aspects pertaining to the novel implementations of the NN-EVP framework are discussed. Section 3 describes the applications of the NN-EVP tool in modeling large viscoplastic deformations with isotropic hardening for both synthetic and experimental data as a function of grain size. A summary of the findings and key contributions concludes the work.

\section{Methods} 
\subsection{Elasto-viscoplasticity constitutive formulation based on dual potential}
 Under the assumption of small strains, elasto-viscoplasticity can be modeled by decomposing the strain into its elastic and viscoplastic parts:
\begin{equation}\label{eq::1}
     \bm{\epsilon}=\bm{\epsilon}^{e}+\bm{\epsilon}^{vp},\\ 
\end{equation}      
 where $\bm{\epsilon}$, $\bm{\epsilon}^{e}$, and $\bm{\epsilon}^{vp}$ are the total strain, elastic strain, and viscoplastic strain, respectively. 

In a similar fashion, the specific free energy $\Psi$ can be decomposed into elastic and viscoplastic contributions, denoted by $\Psi^{e}$ and $\Psi^{vp}$ respectively, as follows:
\begin{equation}\label{eq::4}
     \Psi  = \Psi^{e}( \bm{\epsilon}^{e}) + \Psi^{vp}(r, \bm{\alpha},T ), 
\end{equation}    
    where $r$ and $\bm{\alpha}$ are the thermodynamic variables related to isotropic and kinematic hardening, and $T$ is the temperature. Following the Coleman-Noll procedure \cite{noll1974thermodynamics} the Cauchy stress $\sigma$ and thermodynamic forces $R$ and $X$ can then be defined as:
\begin{equation}\label{eq::5}
     \bm\sigma=\dfrac{\partial\Psi}{\partial{\bm\epsilon^e}}=-\dfrac{\partial\Psi}{\partial{\bm\epsilon^{vp}}},
\end{equation}     
\begin{equation}\label{eq::6}
     R=\dfrac{\partial\Psi}{\partial{r}}
\end{equation} 
\begin{equation}\label{eq::7}
     \mathbf{X}=\dfrac{\partial\Psi}{\partial{\bm\alpha}}.
\end{equation} 
    Under isothermal conditions, the intrinsic potential is written as:
\begin{equation}\label{eq::8}
    \Psi_{I}=\Psi_{I}^{e}(\bm\epsilon^e)+\Psi_{I}^{vp}(\bm\alpha,r). 
\end{equation}     
The generalization of the concept of equipotential surfaces in stress space, representing similar dissipation (i.e., strain rate) levels leads to the definition of a dual potential in the form of \cite{lemaitre1994mechanics}:
\begin{equation}\label{eq::9}
    \varphi^*=\varphi^*(\bm\sigma,R,\mathbf{X};r,\bm\alpha). 
\end{equation} 
The dual dissipation potential $\varphi^*$ satisfies the following conditions:
\begin{itemize}
    \item $\varphi^*$ is convex w.r.t to all its variables, 
    \item $\varphi^*$ is always positive: $\varphi^*>0$, 
    \item $\varphi^*$ includes the origin: $\varphi^*(0,0,0,\bm \alpha,r)=0$.
\end{itemize}
Following the normality law of generalized standard materials the plastic strain rate $\dot{\epsilon^{vp}}$  and the hardening variables $r$ and $\alpha$ can then be obtained as \cite{lemaitre1994mechanics}: 
\begin{equation}\label{eq::10}
  {\dot\epsilon^{vp}}=\dfrac{\partial\varphi^*}{\partial\bm\sigma},
\end{equation}  
\begin{equation}\label{eq::11}
  \dot{r}=-\dfrac{\partial\varphi^*}{\partial{R}},
\end{equation}
\begin{equation}\label{eq::12}
  \dot{\bm\alpha}=-\dfrac{\partial\varphi^*}{\partial\mathbf{X}},
\end{equation}
which leads to the second law of thermodynamics as follows:
\begin{equation}\label{eq::13}
 \Psi_{I}=\bm\sigma:\dfrac{\partial\varphi^*}{\partial\bm\sigma}+\mathbf{X}:\dfrac{\partial\varphi^*}{\partial\mathbf{X}}+R\dfrac{\partial\varphi^*}{\partial{R}} \geq \varphi^*  \geq 0.
\end{equation}
\subsubsection{Decomposition of dual potential}
Under consideration of the recovery effects caused by dislocation annihilation we can formulate the dual potential $\varphi^*$ to also be decomposed into viscoplastic hardening and recovery potentials as follows \cite{lemaitre1994mechanics}:
\begin{equation}\label{eq::14}
  \varphi^*=\Omega_{vp} + \Omega_r,
\end{equation}
where
\begin{equation}\label{eq::15}
    \Omega_{vp}=\Omega_{vp}(\bm\sigma,R,\mathbf{X};r,\bm\alpha),
\end{equation}
\begin{equation}\label{eq::16}
   \Omega_r=\Omega_r(R,\mathbf{X};r,\bm\alpha).
\end{equation}
In the absence of recovery effects, the dual potential can be considered to be equal to the hardening potential $\Omega_p$ \cite{lemaitre1994mechanics}:
\begin{equation}\label{eq::17}
    \varphi^*=\Omega_{vp}=\varphi^*(\bm\sigma,R,\mathbf{X};r,\bm\alpha).
\end{equation}

\subsubsection{Particular functional forms and power-law}
Assuming linear elastic behavior for the elastic part of the free energy function we can write
\begin{equation}
    \Psi_{I}^{e}(\bm{\epsilon}^{e}) = \frac{1}{2} \bm{\epsilon}^{e} : \mathbb{C}:\bm{\epsilon}^{e}
\end{equation}
where $\mathbb{C}$ is the fourth-order elastic stiffness tensor. This yields the Cauchy stress through Hook's law as follows
\begin{equation}\label{eq::2}
           {\bm\sigma}=\mathbb{C}:\bm{\epsilon}^e.
\end{equation}      
For isotropic materials, the stiffness tensor is uniquely defined by the Young's modulus $E$ and Poisson's ratio $\nu$
\begin{equation}\label{eq::stiffness}
    \mathbb{C}_{ijkl} = \frac{E \nu}{(1+\nu) (1-2\nu)} \delta_{ij} \delta_{kl} + \frac{E}{2 (1+ \nu)} (\delta_{ik} \delta_{jl} + \delta_{il} \delta_{jk})
\end{equation}     
    where $\delta_{ij}$ denotes the Kronecker delta. 
    

Under the assumption of isotropic and pressure-independent material response, we can reformulate the dual potential as a function of only the second and third stress invariants
\begin{equation}\label{eq::18}
    \varphi^*=\varphi^*(J_2(\bm\sigma-\mathbf{X}),J_3(\bm\sigma-\mathbf{X}),R,\mathbf{X};r,\bm\alpha),
\end{equation}
where 
\begin{equation}\label{eq::19}
    J_2(\bm\sigma-\mathbf{X})=\sqrt{\dfrac{3}{2}}{\lVert{\bm\sigma^{'}-\mathbf{X}^{'}}\rVert},
\end{equation}
\begin{equation}\label{eq::20}
    J_3(\bm\sigma-\mathbf{X})=\dfrac{1}{3}tr(\bm\sigma-\mathbf{X}),
\end{equation}
and
\begin{equation}\label{eq::21}
    \bm\sigma^{'}=\bm\sigma-\dfrac{1}{3}tr(\bm\sigma),
\end{equation}
\begin{equation}\label{eq::22}
    \mathbf{X}^{'}=\mathbf{X}-\dfrac{1}{3}tr(\mathbf{X}).
\end{equation}
Here, the apostrophe $\square^{'}$ indicates the deviatoric component of the tensor and $tr(\,)$ implies the trace of the argument. 
After assuming that due to the incompressible behavior of metals, only the deviatoric components of the stress contributes to the plastic deformation, the dual potential reduces to \cite{lemaitre1994mechanics}:
\begin{equation}\label{eq::23}
     \varphi^*=\varphi^*(J_2(\bm\sigma-\mathbf{X}),R,\mathbf{X};r,\bm\alpha),
\end{equation}
The fundamental flow rule in viscoplasticity theory involving the thermodynamic force $R(r)$ is commonly written in the form of a power-law and is desirable because it provides uniqueness of solution for the stress value  that accommodates an imposed strain rate \cite{knezevic2016numerical}:
\begin{equation}\label{eq::24}
     \varphi^*=\dfrac{\dot\epsilon_0}{n+1}{\abs{\dfrac{ J_2(\bm\sigma-\mathbf{X})}{R(r)}}^{n+1}}.
\end{equation}
Incorporating Eq. \eqref{eq::24} and Eq. \eqref{eq::19} into Eq. \eqref{eq::10} we get that
\begin{equation}\label{eq::26}
    \dot{\epsilon^{vp}}=\dfrac{3}{2}\dot\epsilon_0{\abs{\dfrac{ J_2(\bm\sigma-\mathbf{X})}{R(r)}}^{n}}\dfrac{\bm\sigma^{'}-\mathbf{X}^{'}}{\lVert{\bm\sigma^{'}-\mathbf{X}^{'}}\rVert}.
\end{equation}

The power-law embeds the introduced strain rate sensitivity and provides uniqueness of solution for the threshold of equivalent von Mises stress accommodating an imposed strain rate. In Eqs \ref{eq::23} and \ref{eq::24}, $n$ is the rate sensitivity parameter and $\dot\epsilon_0$ is a reference strain rate usually chosen as the norm of applied strain rate tensor ${\lVert {\dot{\bm\epsilon}^{app}} \rVert}$. Following Eq. \ref{eq::11}, the variable $r$ and its rate $\dot{r}$, implying the effective accumulated visco-plastic strain and effective visco-plastic strain rate can be written as follows:
\begin{equation}\label{eq::27}
     r=\epsilon_{eff}^{vp}=\int_{0}^{t} \sqrt{\dfrac{2}{3}}\lVert{\dot{\bm\epsilon^{vp}}(\tau)\rVert} d\tau ,
\end{equation}
\begin{equation}\label{eq::28}
     \dot{r}=\dot\epsilon_{eff}^{vp}=\sqrt{\dfrac{2}{3}}\lVert{\dot{\bm\epsilon^{vp}}\rVert}.
\end{equation}

\subsubsection{Perfect viscoplasticity}
\begin{figure}[H]
    \centering
    \includegraphics[scale=0.55]{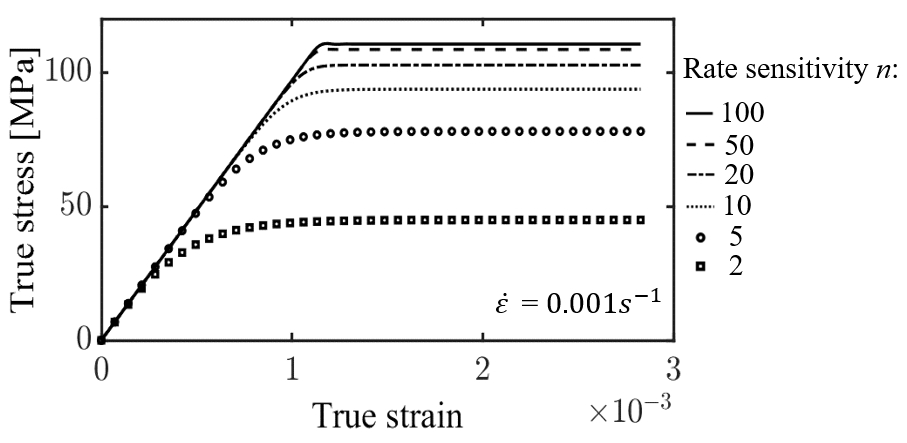}
    \caption{Effect of rate sensitivity on perfect viscoplastic deformation of Cu generated using the power law equation.}
    \label{fig:Perfect_visco_n}
\end{figure}

In case of perfect viscoplasticity and no hardening effects, the yield surface does not evolve as a function of accumulated viscoplastic strain, and the term $R(r)$ reduces to an initial yield value $\sigma^Y$. The dual potential and viscoplastic strain rate can then be written as:
\begin{equation}\label{eq::potential_perfect}
\varphi^*=\dfrac{\dot\epsilon_0}{n+1}{\abs{\dfrac{ \sigma_{eq}}{{\sigma^Y}}}^{n+1}},
\end{equation}
\begin{equation}\label{eq::viscoplastic_strain_rate_perfect}
    \dot{\bm\epsilon^{vp}}=\dfrac{3}{2}\dot\epsilon_0{\abs{\dfrac{ \sigma_{eq}}{\sigma^Y}}^{n}}\dfrac{\bm\sigma^{'}}{\sigma_{eq}},
\end{equation}
where $\sigma_{eq}=\sqrt{\dfrac{3}{2}}{\lVert{\bm\sigma^{'}}\rVert}$. 
Depending on the material properties, the rate sensitivity parameter varies in the range of $10 \leq  n \leq 400$. Figure \ref{fig:Perfect_visco_n} shows the effect of rate sensitivity on perfect viscoplastic flow response of Cu under applied quasi-static strain rate of 0.001$s^{-1}$ generated by the power law equation mentioned above. Note that while higher values of $n$ represent the rate sensitivity of the material more accurately, the computational cost and number of iterations required to numerically solve Eq. \eqref{eq::viscoplastic_strain_rate_perfect} increases. 

\subsubsection{Isotropic hardening} 
\begin{figure}[H]
    \centering
    \includegraphics[scale=0.4]{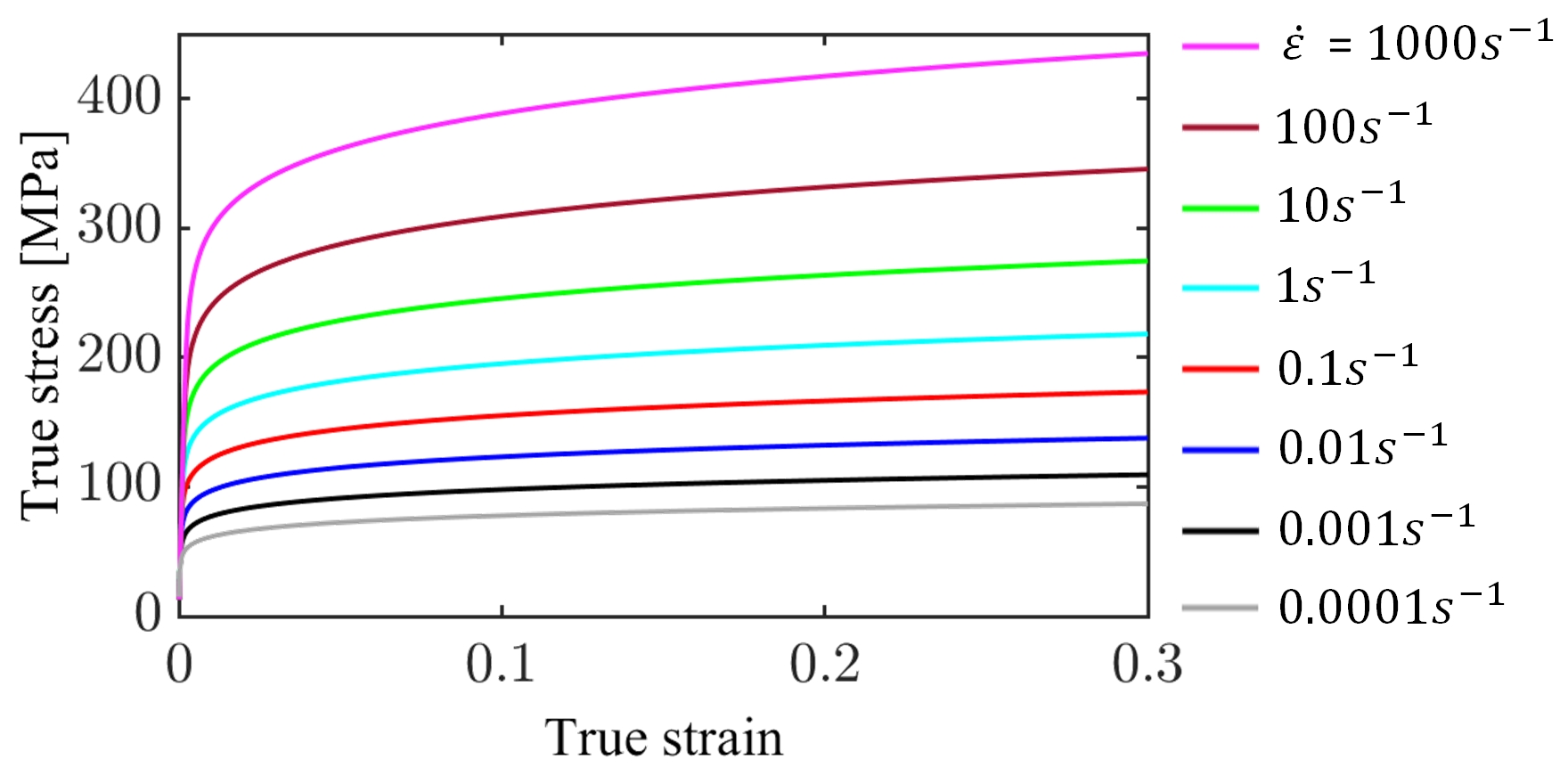}
    \caption{Effect of applied strain rate in viscoplastic flow response of Cu generated by the power law with a rate sensitivity of $n$=20.}
    \label{fig:Rate_effect}
\end{figure}

In the case of isotropic hardening, the yield function $R(r)$ starts with an initial value $\sigma^Y$ ($R(r=0)=\sigma^Y$) and evolves as a function of accumulated viscoplastic strain $r$. The dual potential and viscoplastic strain rate then yield
\begin{equation}\label{eq::potential_hardening}
\varphi^*=\dfrac{\dot\epsilon_0}{n+1}{\abs{\dfrac{ \sigma_{eq}}{R(r)}}^{n+1}},
\end{equation}  
\begin{equation}\label{eq::30}
    \dot{\bm\epsilon^{vp}}=\dfrac{3}{2}\dot\epsilon_0{\abs{\dfrac{ \sigma_{eq}}{R(r)}}^{n}}\dfrac{\bm\sigma^{'}}{\sigma_{eq}}.
\end{equation}
Figure \ref{fig:Rate_effect} shows the viscoplastic flow response in Cu generated by the power law and as a function of applied strain rate and rate sensitivity parameter $n=20$, demonstrating significant anisotropy in the flow response as a function of imposed strain rate.    

While the particular functional forms and the power law model described above enable the constitutive formulation for viscoplastic flow response in metals, predicting the stress-strain behavior using these models often require manual calibration of a set of hardening parameters with experimental data obtained from uniaxial deformation tests. While the automation of the calibration procedures have been enabled by the existing ML and GA models, training these models can be still a time-consuming task, especially when discrepancies of the experimental response and the suggested power law model are present. Therefore, we propose a data-driven elasto-viscoplasticity framework based on neural networks, NN-EVP, that replaces the power law with a generic neural networks based algorithm capable of predicting the flow response in metals and alloys at large deformations in the context of limited data availability.

\subsection{NN-EVP} 
In order to employ a physics-informed data-driven model for the elasto-viscoplastic constitutive formulation described earlier, the dual potential $\varphi^*$ is represented by the response of a neural  ${\mathcal{NN}_{\phi^{*}}}$ whose (scalar) input is chosen based on which hardening phenomena we aim to be captured. This is discussed in more detail in the subsequent sections. 

To remain consistent with thermodynamics laws, the following conditions are (implicitly) enforced on the dual potential neural network ${\mathcal{NN}_{\phi^{*}}}$:

\begin{itemize}
    \item ${\mathcal{NN}_{\phi^{*}}}$ is convex and monotonically increasing 
    \item ${\mathcal{NN}_{\phi^{*}}}$ is always positive: ${\mathcal{NN}_{\phi^{*}}}$ > 0 
    \item ${\mathcal{NN}_{\phi^{*}}}$ includes the origin: ${\mathcal{NN}_{\phi^{*}}(0)}=0$   
\end{itemize}

If ${\mathcal{NN}_{\phi^{*}}}: \mathbb{R} \rightarrow \mathbb{R}$ is a feedforward neural network with $L$ hidden layers, input ${x_0}$ and output ${y_L}$, the neural network can be written as follows:
\begin{equation}
    \begin{aligned}
            x_{0} &\in \mathbb{R}_{\geq 0}, \\
            {x}_{1} =  \mathcal{F}_{1} \left( x_{0} {W}_{1}^{T} + {b}_{1} \right) &\in \mathbb{R}^{n^{1}}, \\
            {x}_{l} = \mathcal{F}_{l} \left( {x}_{l-1} {W}_{l}^{T} + {b}_{l} \right) &\in \mathbb{R}^{n^{l}}, \qquad l=1, \ldots, L-1 \\
            x_{L} = {x}_{L-1} {W}_{L}^{T} + {b}_{L}, &\in \mathbb{R},
    \end{aligned}
\end{equation}
where the weights ${W}_{l}\in \mathbb{R}^{n^{l}\times n^{l-1}}$ and biases ${b}_{l}\in \mathbb{R}^{n^{l}}$ define the set of trainable parameters and the activation functions are denoted by $\mathcal{F}_{1}$. 
The neural network ${\mathcal{NN}_{\phi^{*}}}$ is positive, monotonically increasing, and input convex when the following conditions are met \cite{fuhg2023modular}:
 
\begin{itemize}    
     \item $x_{0}\geq 0 ,  W_{l} \geq 0 , b_{l} \geq 0 $,  
     \item $\rev{\mathcal{F}}_{l}: \mathbb{R}_{\geq 0} \rightarrow \mathbb{R}_{\geq 0}  ,   l=1, \ldots, L$,     
     \item $\rev{\mathcal{F}}_{l}': \mathbb{R}_{\geq 0} \rightarrow \mathbb{R}_{\geq 0}  ,    l=1, \ldots, L $,   
     \item $\rev{\mathcal{F}}_{l}'': \mathbb{R}_{\geq 0} \rightarrow \mathbb{R}_{\geq 0}  ,     l=1, \ldots, L $.  
\end{itemize}
 
In order to satisfy the conditions above, we choose a parameterized and adaptive/scalable Softplus activation function \cite{gnanasambandam2022self,jagtap2022important} as follows: 
 \begin{equation}
     \mathcal{F}_{l}^{\mathcal{NN}_{\phi^{*}}}(x) = \frac{1}{\beta} \log(1+e^x) -\mathcal{F}_{l}^{\mathcal{NN}_{\phi^{*}}}(0) ,
 \end{equation}
 where $\beta>0$ is a training parameter in addition to the weights and biases of the neural network, providing a more generic form of the activation function and thus more flexibility in training. Note that $\beta$ is trainable for each hidden layer of ${\mathcal{NN}_{\phi^{*}}}$. The deduction of $\mathcal{F}_{l}^{\mathcal{NN}_{\phi^{*}}}(0)$ from the equation above is to satisfy the condition ${\mathcal{NN}_{\phi^{*}}(0)}=0$.  Once the dual potential ${\mathcal{NN}_{\phi^{*}}}$ is established, it can be used to replace the particular power law form introduced earlier in Eqs \eqref{eq::potential_perfect} and \eqref{eq::potential_hardening}. Equivalently, the viscoplastic strain rate ${\dot\epsilon^{vp}}$ can be obtained by taking the derivative of the output of ${\mathcal{NN}_{\phi^{*}}}$ with respect to Cauchy stress:
\begin{equation}
   \bm{\dot\epsilon^{vp}}= \dfrac{\partial{\mathcal{NN}_{\phi^{*}} }}{\partial{\bm\sigma}}.
\end{equation}
 
\subsubsection{NN-EVP for perfect viscoplasticity} 
\begin{figure}[H]
    \centering
    \includegraphics[scale=0.3]{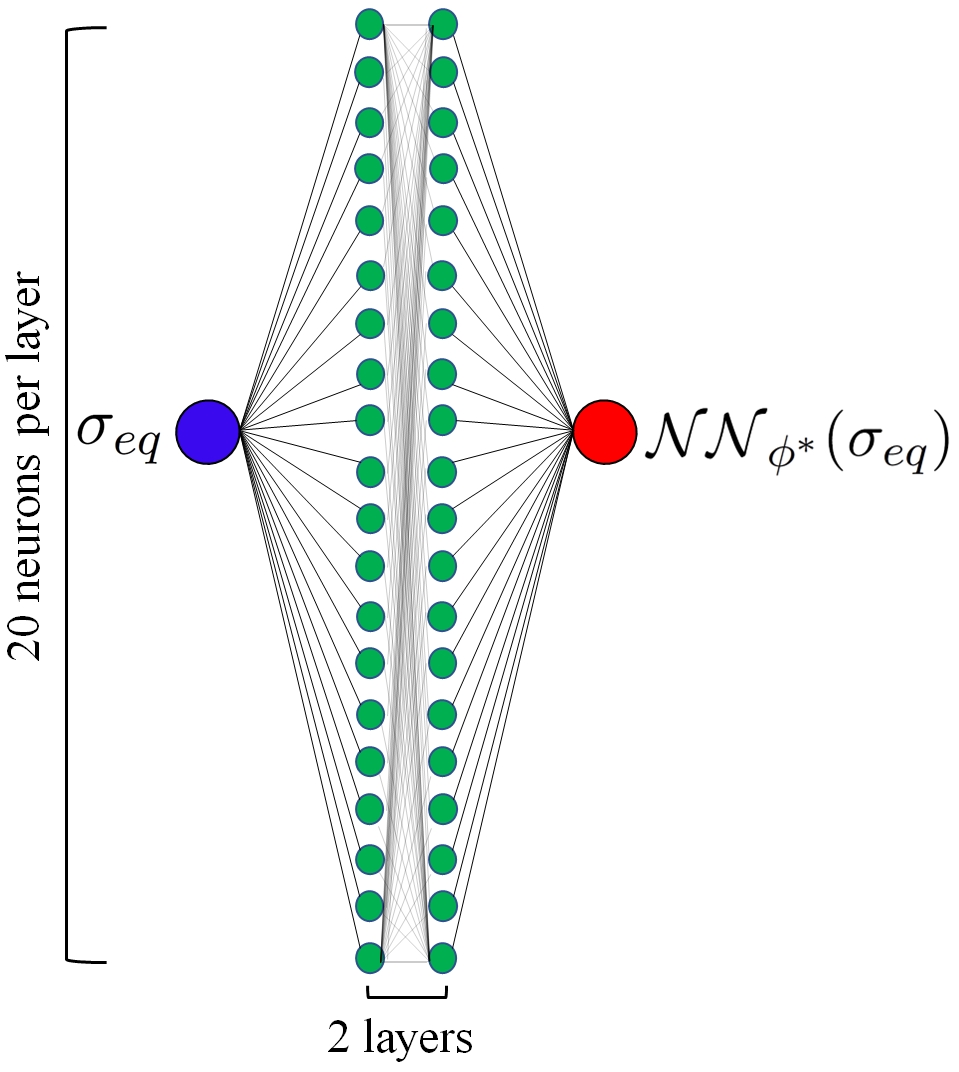}
    \caption{A schematic of the neural network used for perfect elasto-viscoplasticity. }
    \label{fig::ICNN_architecture}
\end{figure}

In the case of perfect viscoplasticity, the yield function $\sigma^Y$ is constant and does not evolve. We can therefore use $\sigma_{eq}$ as the input to the neural network, ${\mathcal{NN}_{\phi^{*}}} ={\mathcal{NN}_{\phi^{*}} (\sigma_{eq}) }$. Figure \ref{fig::ICNN_architecture} shows a schematic of the particular neural network architecture used for modeling perfect viscoplasticity in this work. While different configurations are possible in particular with regards to the selection of the number of layers and neurons, due to the already low amount of training data (less than 100 data points depending on the resolution of stress-strain data points) compared to the number of trainable parameters we show the results for a network with 2 layers and 20 neurons in the following. We remark that slight changes to this network architecture, i.e. 2 layers with 30 neurons, for example, have not shown to yield different results.
\subsubsection{NN-EVP with isotropic hardening} 
\begin{figure}[H]
    \centering
    \includegraphics[scale=0.3]{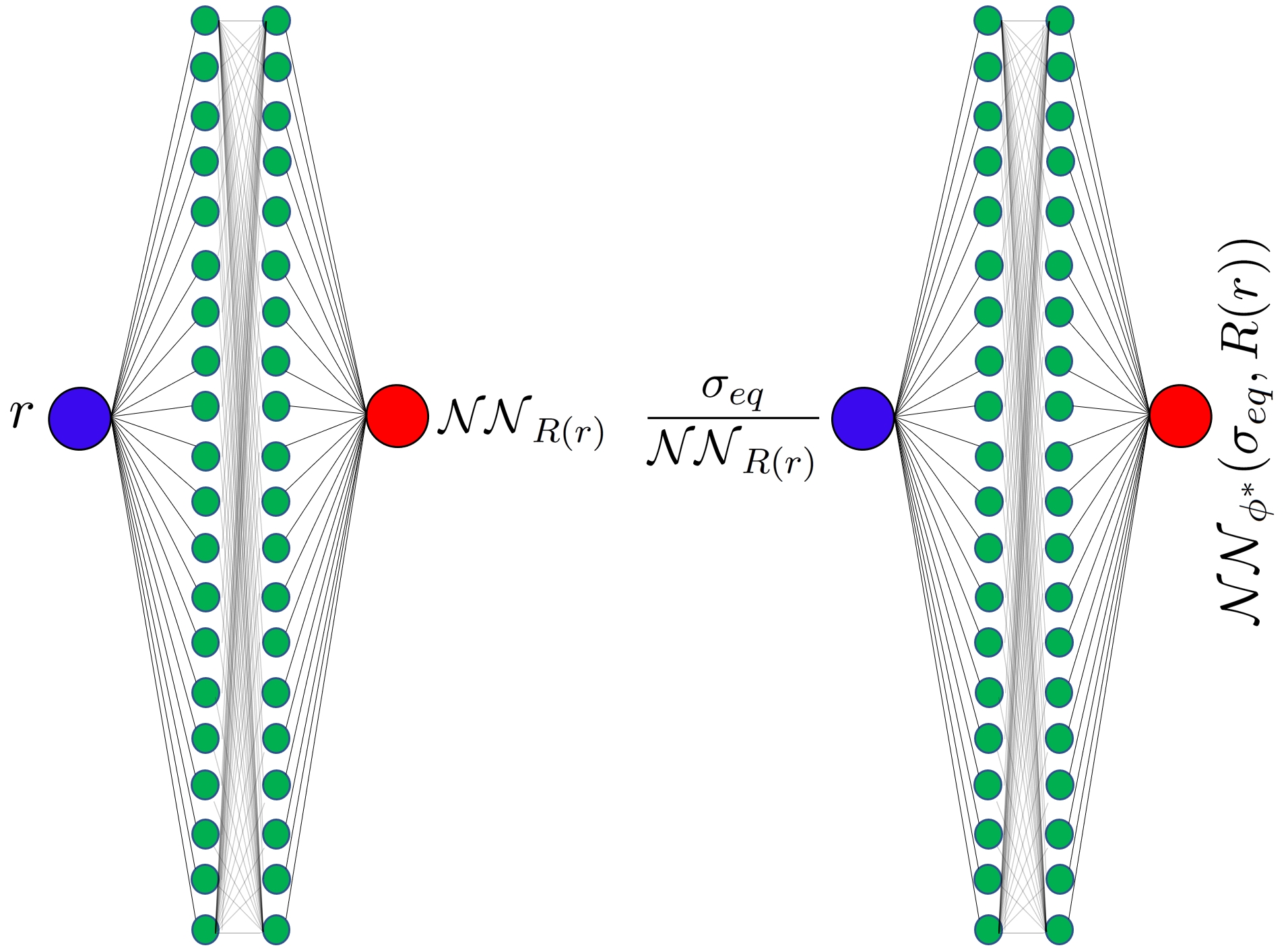}
    \caption{A schematic of the neural network used for elasto-viscoplasticity with isotropic hardening. }
    \label{fig:ICNN_architecture_isotropic_hardening}
\end{figure}

Once hardening effects are introduced into the constitutive model, the yield function ${R(r)}$ (which will be not imposed, but discovered) evolves during the deformation as a function of accumulated plastic strain ${r}$. As a result, information pertaining to the evolution of yield function is also required as an additional input to the dual potential neural network $\mathcal{NN}_{\phi^{*}}$. Thus, we could define the dual potential neural network as $\mathcal{NN}_{\phi^{*}}$ =${\mathcal{NN}_{\phi^{*}} (\sigma_{eq},{R(r)}) }$. To this end, an additional neural network ${\mathcal{NN}_{R}=\mathcal{NN}_{R}(r)}$ is required to track the evolution of  strain hardening. Notice that the output of hardening neural network ${\mathcal{NN}_{R}(r)}$ could be used as an input to the dual potential neural network ${\mathcal{NN}_{\phi^{*}} (\sigma_{eq},{R(r)}) }$. Here, instead of taking $\sigma_{eq}$ and ${R(r)}$ as two independent inputs to ${\mathcal{NN}_{\phi^{*}}}$, we choose to take the ratio $\dfrac{\sigma_{eq}}{{R(r)}}$ as a single input. This mechanistic assumption informs the neural network with the physical constraint that $\sigma_{eq}$ needs to be scaled proportionally with the evolution of $R(r)$, imposing the strain hardening effects as the ratio $\dfrac{\sigma_{eq}}{{R(r)}}$ controls the elasto-viscoplastic transition as well as the rate of the evolution of viscoplastic strain rate. Figure \ref{fig:ICNN_architecture_isotropic_hardening} shows a schematic of NN-EVP with isotropic hardening. To remain consistent with the thermodynamics laws, the following conditions are enforced on the hardening neural network $\mathcal{NN}_{R}$:

\begin{itemize}  
    \item $\mathcal{NN}_{R}$ is monotonically increasing,   
    \item $\mathcal{NN}_{R}$ is always positive: ${\mathcal{NN}_{R}} > 0 $,  
    \item $\mathcal{NN}_{R}$ does not include the origin: $\mathcal{NN}_{R}(r=0) \neq 0$.   

\end{itemize}
The neural network $\mathcal{NN}_{R}$ is positive and monotonically increasing when the following conditions are met:
\begin{itemize}    
     \item $x_{0}\geq 0 ,  W_{l} \geq 0, b_{l} \geq 0 $,  
     \item $\rev{\mathcal{F}}_{l}: \mathbb{R}_{\geq 0} \rightarrow \mathbb{R}_{\geq 0}  ,   l=1, \ldots, L$,    
     \item $\rev{\mathcal{F}}_{l}': \mathbb{R}_{\geq 0} \rightarrow \mathbb{R}_{\geq 0}  ,    l=1, \ldots, L $.  
\end{itemize}
Notice that in contrast to the dual potential network $\mathcal{NN}_{\phi^{*}}$, the hardening network $\mathcal{NN}_{R}$ does not contain the origin and does not require convexity. This is due to the fact that the yield function ${R(r)}$ has a nonzero value equal to the initial yield at zero accumulated viscoplastic strain $r$. We also remark that since $\mathcal{NN}_{R}$ is monotonically increasing, the reciprocal form of $\dfrac{1}{\mathcal{NN}_{R}}$, used as input to $\mathcal{NN}_{\phi^{*}}$, is monotonically decreasing \cite{fuhg2023modular}.  
In order to satisfy the conditions above, we choose combinations of forms of ReLU, adaptive logistic or adaptive tanh activation functions as follows: 
 \begin{equation}\label{eq:ReLU_logistic}
     \rev{\mathcal{F}_{l}}^{\mathcal{NN}_{R}}(x) = \alpha_1 \max(x,0) + \alpha_2  \frac{1}{1+e^ {-\beta x}},
  \end{equation}
  \begin{equation}\label{eq:ReLU_tanh}    
     \rev{\mathcal{F}_{l}}^{\mathcal{NN}_{R}}(x) = \alpha_1 \max(x,0) + \alpha_2 \frac{e^{\beta  x}-e^{- \beta x}}{e^{\beta x}+e^{-\beta x}},
 \end{equation}
   \begin{equation}\label{eq:Logistic_tanh}    
     \rev{\mathcal{F}_{l}}^{\mathcal{NN}_{R}}(x) = \alpha_1 \frac{1}{1+e^ {-\beta x}} + \alpha_2 \frac{e^{\beta  x}-e^{- \beta x}}{e^{\beta x}+e^{-\beta x}},
 \end{equation}
 where $\alpha_1$ and $\alpha_2$ denote the weights assigned to the activation functions and applied to the hidden layers of each neural network. The reason behind these mixed activation functions is to facilitate learning the hardening response for larger strain amplitudes. Since the logistic activation function saturates early at lower strain levels, the addition of either a ReLu or tanh response compensates for stress-strain curvature at later stages of hardening. The effect of variation in $\alpha_1$ and $\alpha_2$  as well as the particular selection of the activation functions on the flow response is discussed later in detail. 
\subsubsection{Hall-Petch effects and grain size-aware NN-EVP}
Hall-Petch effects and variation in grain size are concomitant with strong anisotropy in viscoplastic flow responses. In particular, the initial yield function of the material alters drastically depending on the average grain size of the underlying microstructure. Particular functional forms for Hall-Petch relationship vary in the literature. However, the most common form  \cite{hansen2004hall,saada2005hall,cordero2016six,eghtesad2022density,eghtesad2023machine} can be written as follows:
\begin{equation}\label{eq::Hall_Petch}
R_{0,HP}=\dfrac{H\mu\sqrt{b} }{\sqrt{d_{grain}} }    
\end{equation}
where $b$ is the Burgers vector, $\mu$ is the shear modulus, $d_{grain}$ is the average grain size and $H$ stands for the Hall-Petch coefficient usually obtained from calibrations with experimental observations. Notice that Eq. \eqref{eq::Hall_Petch} provides a nonlinear relationship between the grain size and initial yield function in non-logarithmic  space, however, it can be shown that the Hall-Petch stress does not always scale as the inverse square root of grain size \cite{dunstan2014grain} and thus Eq. \ref{eq::Hall_Petch} is simply a specific function of many possible forms. In order to incorporate the Hall-Petch effect into the NN-EVP framework, we introduce a Hall-Petch neural network $\mathcal{NN}_{HP}=\mathcal{NN}_{HP}({d_{grain}})$ in addition to the dual potential $\mathcal{NN}_{\phi^{*}}$ and the hardening neural network $\mathcal{NN}_{R}$. This allows us to learn and discover the Hall-Petch relationship. To remain consistent with thermodynamics laws, the following conditions are enforced on the Hall-Petch neural network $\mathcal{NN}_{HP}$:

\begin{itemize}
    \item $\mathcal{NN}_{HP}$ is monotonically decreasing, 
    \item $\mathcal{NN}_{HP}$ is always positive: ${\mathcal{NN}_{HP}}$ > 0, 
    \item $\mathcal{NN}_{HP}({d_{grain} \rightarrow 0}) \rightarrow \infty$.
\end{itemize}
The neural network $\mathcal{NN}_{HP}$ is positive when the following conditions are met:
\begin{itemize}    
     \item $x_{0}\geq 0 ,  W_{l} \geq 0, b_{l} \geq 0 $,  
     \item $\rev{\mathcal{F}}_{l}: \mathbb{R}_{\geq 0} \rightarrow \mathbb{R}_{\geq 0}  ,   l=1, \ldots, L$.     
\end{itemize}
Here, we first choose a standard tanh activation function as follows and then take the reciprocal of the network output to satisfy the conditions above: 
 \begin{equation}
     \rev{\mathcal{F}_{l}}^{\mathcal{NN}_{{HP}}(x)} = \frac{e^{x}-e^{-x}}{e^{x}+e^{-x}}, 
 \end{equation}
 \begin{equation}
     \mathcal{NN}_{HP} \leftarrow \frac{1}{\mathcal{NN}_{HP}}. 
 \end{equation}
 
\begin{figure}[H]
    \centering
    \includegraphics[scale=0.425]{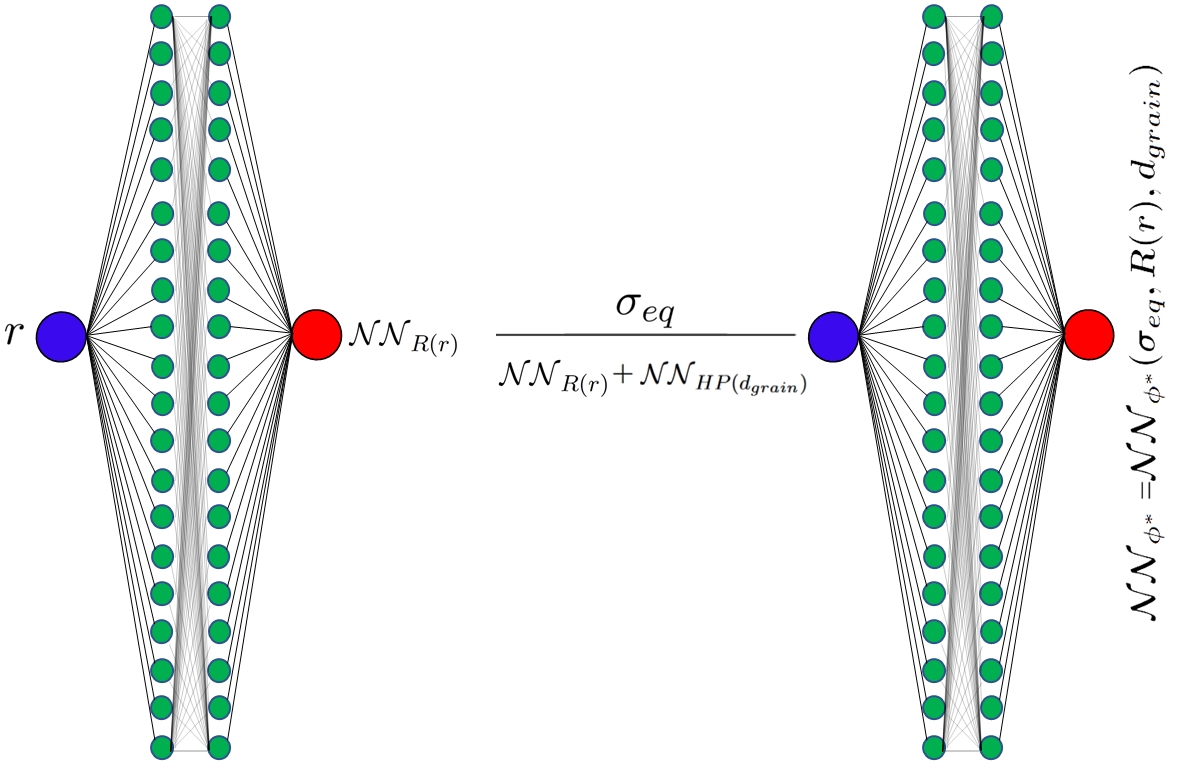}
    \caption{A schematic of neural network for elasto-viscoplasticity with isotropic hardening and Hall-Petch effects.}
    \label{fig:ICNN_architecture_after_HallPetchNN}
\end{figure}

 Notice that since the Hall-Petch term only affects the initial yield function and does not evolve during the deformation, an adaptive activation function here is not necessary. Figure \ref{fig:ICNN_architecture_after_HallPetchNN} shows a schematic of grain size-aware NN-EVP architecture including the Hall-Petch effects. As illustrated in the figure, the response of the Hall-Petch network $\mathcal{NN}_{HP}$ and the hardening network $\mathcal{NN}_{R}$ are combined to $\dfrac{\sigma_{eq}}{{\mathcal{NN}_{R(r)}+\mathcal{NN}_{HP}(d_{grain})}}$ and then used as an input to the dual potential network $\mathcal{NN}_{\phi^*}$.

\section{Results and Discussion} 
\subsection{Implementation highlights} 
The NN-EVP framework is implemented in PyTorch \cite{paszke2019pytorch}. PyTorch takes advantage of automatic differentiation \cite{paszke2017automatic} facilitating automatic computation of local gradients of the output of neural network with respect to its inputs. Automatic differentiation also allows us to determine the Jacobian required for solving the update step of implicit time-stepping using the Newton Raphson (NR) method \cite{ypma1995historical}. Details of the implementation are provided in the form of a pseudo-code in algorithm \ref{alg:Train}. Due to its adaptive learning decay algorithm and robustness, the AdamW optimizer \cite{loshchilov2017decoupled} is utilized to optimize the parameters of the three neural network used in this study. In addition, a cosine annealing scheduler \cite{loshchilov2016sgdr,lu2020multi} is utilized to further enhance the  learning rate adaptivity and therefore improve the robustness of the NN-EVP framework. The cosine annealing scheduler is set with a starting learning rate of $1e^{-2}$ and saturates to a learning rate of $1e^{-3}$. 

The established framework is then trained to model perfect viscoplasticity as well as an isotropic hardening response under uniaxial tensile loading. We consider two different scenarios. First, the flow response is synthetically generated using the power law and phenomenological Johnson-Cook hardening \cite{johnson1983constitutive} where in addition to fitting the existing data, the flow response is predicted to larger strain amplitudes via extrapolations enabled by recovering the trained neural networks. Next, experimentally measured data pertaining to large deformations as a function of grain size are used to train the NN-EVP framework presented herein. In all cases, a constant 11 component of the strain rate $\dot\epsilon^{app}=1e^{-3} s^{-1}$ in X direction is applied. In order to satisfy the uniaxial loading conditions, the 22 and 33 components of the stress tensor are enforced to be zero (see algorithm \ref{alg:Train}). The NR solver tolerance is set to $\mathcal{TOL_{NR}}=1e^{-6}$ for all simulations which is reached within 2-4 iterations depending on the deformation history. Finally, the mean squared error function is used to calculate the loss value in each training epoch.      
\begin{algorithm}
\caption{Grain size aware NN-EVP}\label{alg:Train}
\begin{algorithmic}   
\Require Stress-strain response obtained from power law or experimental data:  
                $\begin{bmatrix}
    \bm\epsilon^{True} & , & \bm\sigma^{True} \\
               \end{bmatrix}$

\Establish 
\begin{itemize}
    \item Initialize positive, monotonically increasing, input convex neural network $\mathcal{NN_{\phi^*}}$  and its corresponding optimizer $\mathcal{O_{NN_{\phi^*}}}$ and parameters $\mathcal{\theta_{NN_{\phi^*}}}$,
    \item Initialize positive, monotonically decreasing neural network $\mathcal{NN_{R}}$  and its corresponding optimizer $\mathcal{O_{NN_{R}}}$ and parameters $\mathcal{\theta_{NN_{R}}}$,
   \item Initialize positive, monotonically decreasing neural network $\mathcal{NN_{HP}}$  and its corresponding optimizer $\mathcal{O_{NN_{HP}}}$  and parameters $\mathcal{\theta_{NN_{HP}}}$ ,
    \item Define applied strain rate tensor corresponding to the uniaxial deformation : 
    $\begin{bmatrix}
    \dot\epsilon^{app} & 0 & 0\\
     0 & -\dfrac{\dot\epsilon^{app}}{2} & 0 \\
     0 & 0 & -\dfrac{\dot\epsilon^{app}}{2}  
\end{bmatrix}$,
    \item Define total number of time steps $N_{Steps}$ , training epochs $N_{Epochs}$ and NR tolerance $\mathcal{TOL_{NR}}$,  
    \item Define the material properties, Young's modulus $E$ and Poisson's ratio $\nu$ as well as the average grain size $d_{grain}$. 
\end{itemize}
\Begin \\
\textbf{Calculate the elastic stiffness} $\mathbb{C}$: \vspace{1mm}

    $\mathbb{C}_{ijkl} = \frac{E \nu}{(1+\nu) (1-2\nu)} \delta_{ij} \delta_{kl} + \frac{E}{2 (1+ \nu)} (\delta_{ik} \delta_{jl} + \delta_{il} \delta_{jk})$. \vspace{1mm}

\textbf{Initialize strain and stress tensors:} $\bm{\epsilon}^{t=0} = \bm{0}$,  $\bm{\epsilon}^{vp,t=0} = \bm{0}$ , $\bm{\sigma}^{t=0} = \bm{0}$  \vspace{1mm}
\For {$epoch =0, N_{epochs}$} \vspace{1mm}

\textbf{Initialize loss:} $\mathcal{L}(\mathcal{O_{NN_{\phi^*}}},\mathcal{O_{NN_{R}}},\mathcal{O_{NN_{HP}}}) = 0$ \vspace{1mm}
\For {$t =0, N_{steps}$} \vspace{1mm}
  
 \quad  Update the total strain at current time step $\bm{\epsilon}^{t+1}$ using applied strain rate: \vspace{1mm}

 \quad  $\bm{\epsilon}^{t+1} \leftarrow \bm{\epsilon}^{t} + \bm{\dot{\epsilon}}\Delta t $ \vspace{1mm}

 \quad  Calculate trial stress using elastic strain:  \vspace{1mm}

 \quad  $\bm\sigma^{t+1} = \mathbb{C}:\bm{\epsilon}^{t+1} $ \vspace{1mm}
 
 \quad  Calculate viscoplastic strain rate using dual potential neural network $\mathcal{NN_{\phi^*}}$: \vspace{1mm}
 
 \quad  ${\dot{\bm\epsilon}^{vp,t+\Delta t}}= \dfrac{\partial{\mathcal{NN}_{\phi^{*}} (\bm\sigma^{t+1},R^{t+1},d_{grain}) }}{\partial{\bm\sigma}}$ \vspace{1mm} \Comment{Using Torch.autograd.grad()}
  
 \quad  \textbf{Define the residuals}:  \Comment{Use Voigt notation $\sigma_{ij}\rightarrow\sigma_{k},\mathbb{C}_{ijkl}\rightarrow\mathbb{C}_{kl}, k,l=1,6$} \vspace{1mm}
 \begin{fleqn}[\dimexpr(\leftmargini-\labelsep)*2]
          \setlength\belowdisplayskip{0pt}
 \begin{equation*}
 \mathcal{X}_k=\begin{bmatrix}
     {\mathcal{R}es_k(\sigma^{t+1})}=\sigma_k^{t+1}-\mathbb{C}_{kl}({\epsilon_l}^{t+\Delta t}-{\epsilon_l}^{vp,t}-{\dot\epsilon_l}^{vp,t+\Delta t}\Delta t)=0 \\
       {\mathcal{R}es_k(\sigma^{t+1})}=\sigma^{t+1}_{k}  = 0 \, \qquad \text{for } k \neq 1 \\      
               \end{bmatrix}
           \end{equation*}
           \end{fleqn}\\
   \vspace{1mm}
  \qquad \quad \textbf{Deploy NR with Jacobian $\mathcal{J}$ to solve for stress:}  \Comment{NR iterations denoted by $^i$} 
  \While {$\norm{\mathcal{X}} < \mathcal{TOL_{NR}}$}   
              \begin{fleqn}[\dimexpr(\leftmargini-\labelsep)*2] 
              \setlength\belowdisplayskip{0pt}
 \begin{equation*}
                \begin{aligned}[t]
 \sigma_k^{t+1,i+1} = \sigma_k^{t+1,i} - \mathcal{J}_{kl}^{-1} {\mathcal{X}_l(\sigma^{t+1,i})}, \mathcal{J}_{kl}= \dfrac{\partial{\mathcal{X}_k}}{\partial \sigma_l}|_{\sigma^{t+1,i}}           
              \end{aligned}
          \end{equation*}
          \end{fleqn}
\EndWhile  \Comment{End of loop for NR solver} \vspace{1mm}

\textbf{Update optimization loss:} \vspace{1mm}

$\mathcal{L} \leftarrow \mathcal{L} + \norm{ \bm\sigma^{t+1} -  \bm\sigma^{True,t+1}}^{2}$ \vspace{1mm}

\EndFor \Comment{End of loop for time increments} \vspace{1mm}

\textbf{Calculate gradients}: $\bm{G}_{\mathcal{NN}_\phi^*} = \nabla_{{\theta_{\phi^*}} } \mathcal{L}, \bm{G}_{\mathcal{R}} = \nabla_{{\theta_{R}}} \mathcal{L} , \bm{G}_{\mathcal{HP}} = \nabla_{{\theta_{HP}} }\mathcal{L} $  \vspace{1mm}

\textbf{Update NN parameters}: $\mathcal{O}_{R}, \mathcal{O}_{\phi}, \mathcal{O}_{M}$  \Comment{Use gradient clipping enforce positive outputs} \vspace{1mm}

\EndFor \Comment{End of loop for training epochs} \vspace{1mm}
\End
\end{algorithmic}
\end{algorithm}

\subsection{Rediscovering the power law for perfect viscoplasticity} 
\begin{figure}[H]
    \centering
    \includegraphics[scale=0.575]{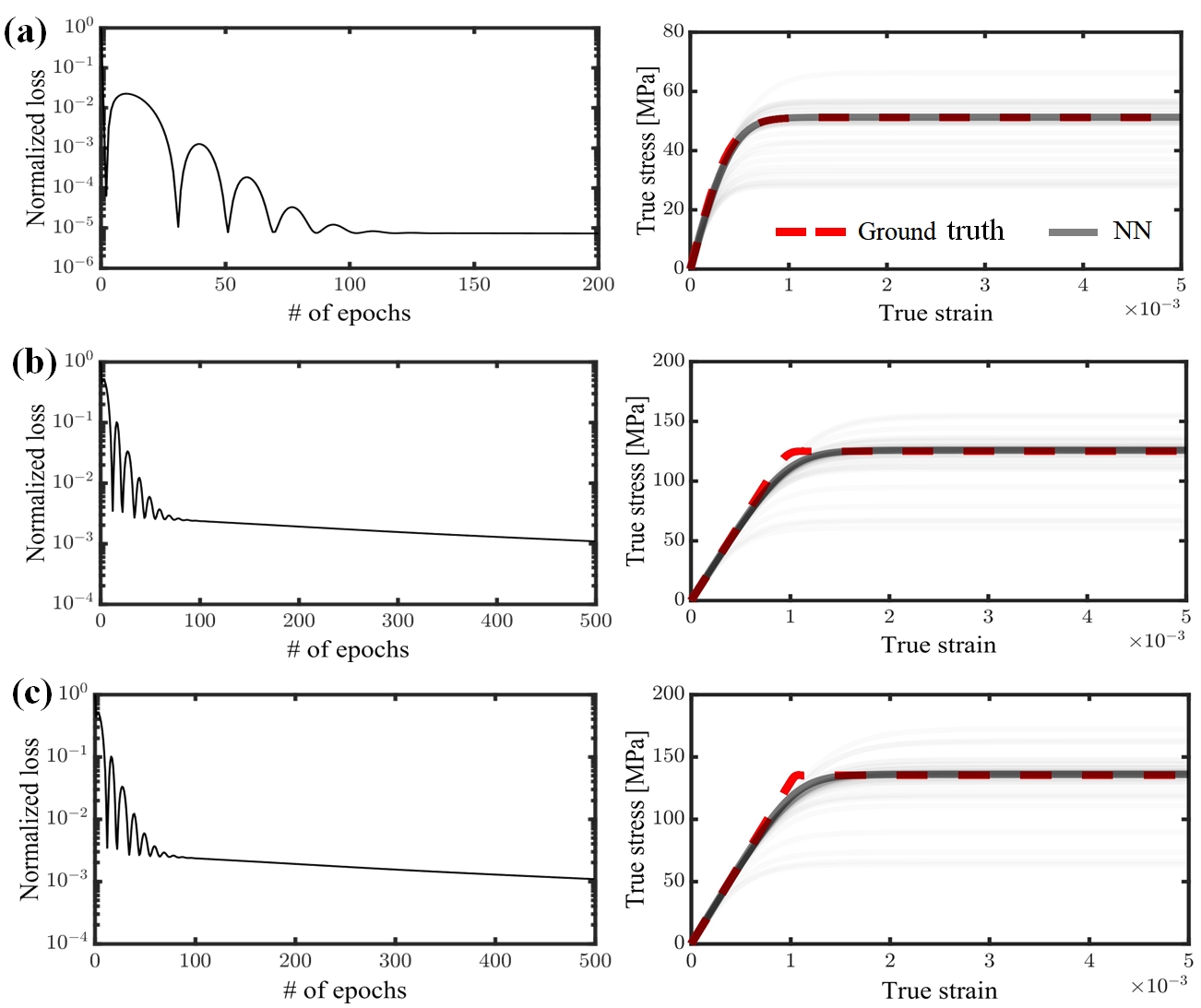}
    \caption{Model fit for perfect elasto-viscoplasticity with power-law rate sensitivities of (a) n=10, (b) n=20 and (c) n=100. Transparent lines indicate the incremental training history.}
    \label{fig:Perfect_visco_flow}
\end{figure}

In order to predict perfect viscoplasticity in Cu, the power law equation with a constant yield function $\sigma^Y$ and rate sensitivities of $n=$10, 20, and 100 are used to generate the synthetic stress-strain data. Since strain hardening is not involved, only the dual potential neural network ${\mathcal{NN}_{\phi^{*}} (\sigma_{eq}=\sigma) }$ is utilized. The left plot of Fig. \ref{fig:Perfect_visco_flow}(a) shows the evolution of normalized loss over 200 training epochs for a rate sensitivity of $n=$10. The training error saturates to a value of $1e^{-5}$ after around 100 iterations. Note that the fluctuations observed in the loss evolution are innate to the optimizer algorithm regardless of the learning rate value. The right plot of Fig. \ref{fig:Perfect_visco_flow}(a) illustrates the trained model output using the ground truth provided by the power law. The transparent lines represent the training history of the model during the optimization process. Figures \ref{fig:Perfect_visco_flow}(b) and \ref{fig:Perfect_visco_flow}(c) highlight the loss evolution and training response for power laws with rate sensitives of $n=$20 and $n=$100 respectively. As the rate sensitivity increases, a sharper transition from elastic to viscoplastic deformation is observed, making it more challenging for the neural network to fit the data at the transition regime. An increased number of training epochs from 200 to 500 is an indication of such behavior. Note that while the training loss is still reducing slowly after around 500 epochs, to maintain consistency and computational efficiency, we hereafter set the maximum number of epochs $N_{Epochs}$=500 for all the simulations.   

\subsection{Training via phenomenological hardening} 
\begin{figure}[H]
    \centering
    \includegraphics[scale=0.475]{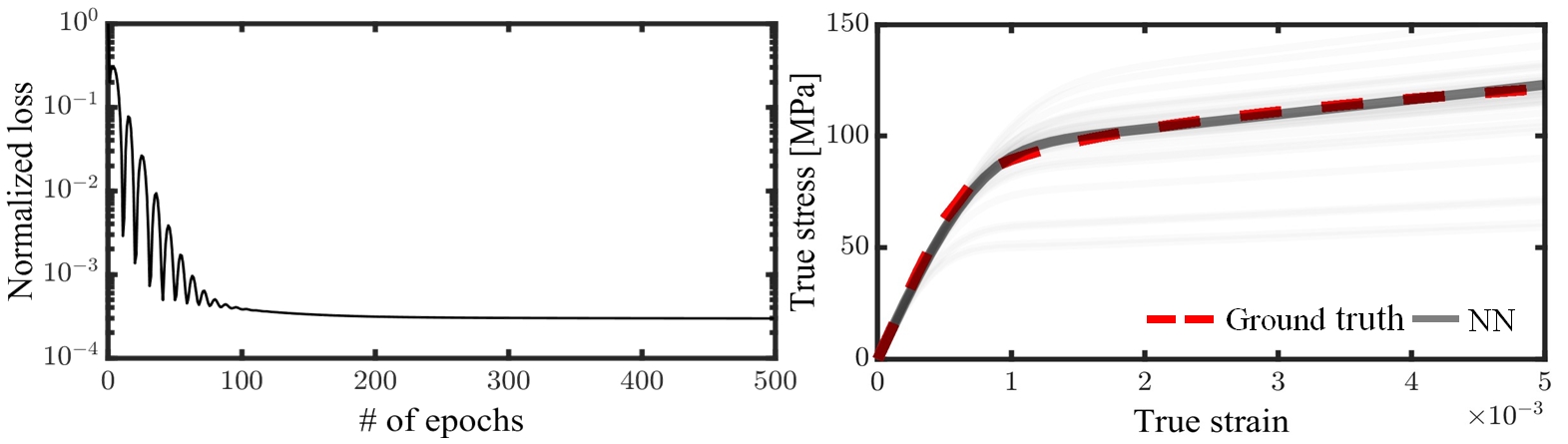}
    \caption{Model fit for elasto-viscoplasticity with isotropic Johnson-Cook hardening. Transparent lines indicate the incremental training history.}
    \label{fig:hardening_visco_flow}
 \end{figure}

\vspace{0mm}
Synthetic stress-strain data for isotropic hardening can be generated using either physics-based or phenomenological hardening models. Physics-based hardening laws are more suited for mesoscale or micromechanical simulations because they consider the evolution of dislocation substructures and dislocation densities \cite{eghtesad2022density} and often require an evolved implementation framework which is computationally intensive. On the other hand, phenomenological hardening models such as Voce \cite{voce1948relationship}, Peirce, Asaro, and Needleman (PAN) \cite{peirce1982analysis} or Johnson-Cook \cite{johnson1983constitutive} are simpler in implementation and thus computationally more efficient for modeling macroscopic behavior where the details pertaining to the underlying microstructure is not considered. In the present work, we use the Johnson-Cook isotropic hardening model with the functional form written as follows \cite{johnson1983constitutive}:
\begin{equation}
    R(r)=[A+Br^n][1+C \log\dfrac{r}{r^*}][1-{T^*}^m],
\end{equation}
\begin{equation}
    T^*=\dfrac{T-T_0}{T_{m}-T_0}, 
\end{equation}
where $A$, $B$, $C$, $n$ and $m$ are the hardening parameters associated with the Johnson-Cook model. Note that since we are considering deformations at room temperature $T_0$, the term $T^*$ that includes the effects of melting temperature $T_m$ vanishes. The values corresponding to the Johnson-Cook model parameters for Cu are listed in Table \ref{tab:JohnsonCookParams} \cite{johnson1983constitutive}. Here both the dual potential network ${\mathcal{NN}_{\phi^{*}} (\sigma, R(r)) }$ and hardening neural network ${\mathcal{NN}_{R(r)}}$ are utilized for training. The dual potential network uses the SoftPlus activation function while the hardening neural network ${\mathcal{NN}_{R(r)}}$ uses the adaptive logistic activation function obtained by setting $\alpha_1=$0 and $\alpha_2=$1 in Eq. \ref{eq:ReLU_logistic}. Figure \ref{fig:hardening_visco_flow} shows the loss evolution and trained model. Young's modulus and Poisson's ratio of 130 GPa and 0.34 are used for the elastic response. The loss evolution experiences an abrupt drop to a normalized value of around $5e^{-4}$ and saturates after around 100 epochs, indicating a fast convergence of the proposed framework.   

\begin{table}
\begin{center}
\begin{tabular}{c c c c c} 
 \hline
 $A[MPa]$  & $B[MPa]$ & $C$ & $m$ & $n$\\ [0.5ex]  
 \hline  
 90 & 292 & 0.31 & 0.025 & 1.09 \\ [0ex] 
 \hline
\end{tabular}
\end{center}
\caption{Johnson-Cook hardening model parameters for Cu.}\label{tab:JohnsonCookParams}
\end{table}

\subsubsection{Extrapolation of flow response to larger strain amplitudes} 
\begin{figure}[H]
    \centering
    \includegraphics[scale=0.95]{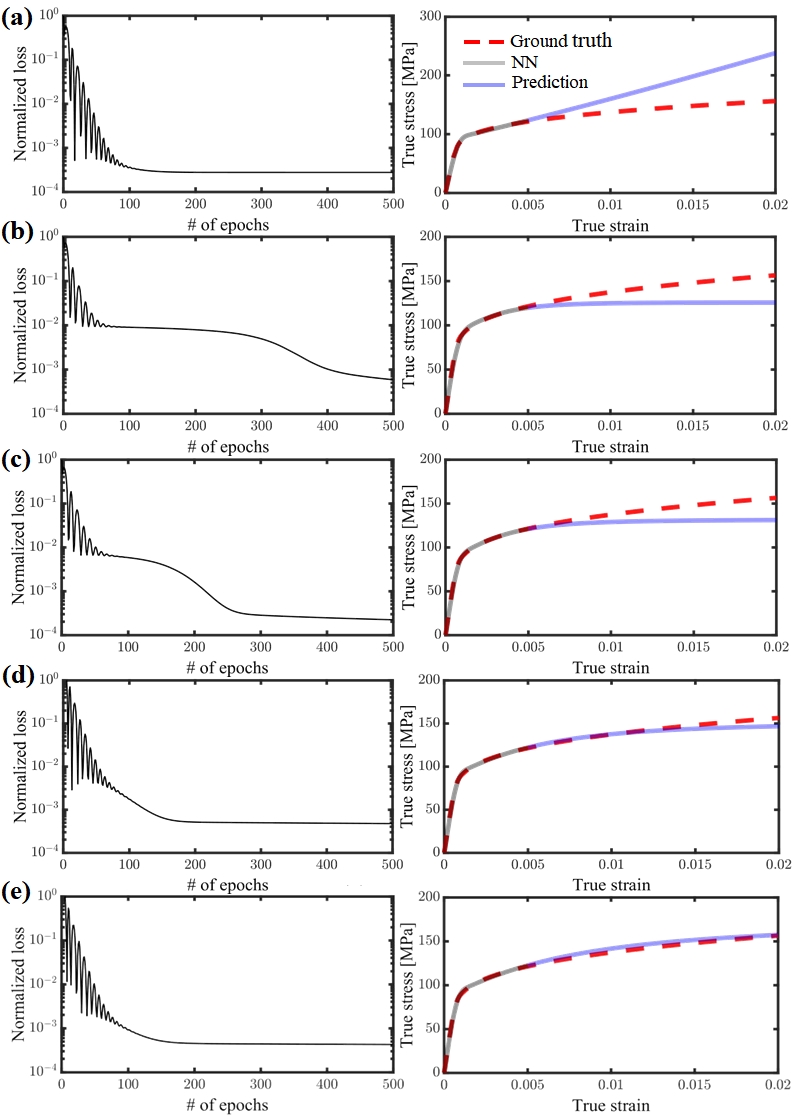}
    \caption{Extrapolations of flow response with an isotropic hardening response via phenomenological Johnson-Cook model as a function of selected activation functions and their combination weights $\alpha_1$ and $\alpha_2$ with (a) 20\% logistic - 80\% ReLU,  (b) 80\% logistic - 20\% tanh, (c) 50\% logistic - 50\% tanh, (d) 100\% tanh and (e) 80\% tanh - 20\% ReLU. The dashed red line shows the ground truth. The solid grey line indicates the trained NN model and the solid blue line implies the extrapolations of the flow response using the NN model.}
    \label{fig:Extrapolations}
\end{figure}

One of the powerful advantages (and challenges) of algorithms that are developed via neural network is the ability to extrapolate beyond the data available to the networks during the training process. Once a model is trained for a specific set of data, the recovered trained model is able to predict the behavior for any given input within the trained framework and beyond the data observed by the model. Here, we restrict the model during training for predictions up to 0.5\% total strain and recover the model and its trained parameters to examine the ability of the model to extrapolate the flow response up to 2\% total strain. This benchmark provides us with a train-test configuration consisting of 25\% training and 75\% testing data. Figure \ref{fig:Extrapolations} shows the extrapolations of stress-strain response for different ratios of the mixed activation functions ReLU, adaptive logistic, and adaptive tanh with variable combination weights $\alpha_1$ and $\alpha_2$ in Eqs. \eqref{eq:ReLU_logistic}, \eqref{eq:ReLU_tanh} and \eqref{eq:Logistic_tanh}.     
The logistic activation function saturates rapidly at low strain amplitudes, hindering the model from properly capturing the hardening response at higher strain amplitudes. To address this issue, tanh, and ReLU activation functions are combined with logistic activation functions using different weights to predict the flow response beyond the training domain. Note that the addition of the ReLU activation function should be done with caution to avoid excessive hardening behavior shown in Fig. \ref{fig:Extrapolations}(a). However, it is worthwhile to mention that sudden increase in strain hardening is well observed in case of deformation twinning in compression tests \cite{salem2003strain} and thus, a mixed activation function with large ReLU weights seems to be a promising approach for capturing the rapid hardening behavior upon twinning formations. Among various configurations shown in Fig. \ref{fig:Extrapolations}(a)-(e), a mixed activation function with 80\% adaptive tanh and 20\% ReLU best predicts the hardening behavior up to 2\% strain.

\subsection{Training large plastic deformations via experimental data}  

So far the capability of the NN-EVP framework in training and extrapolating the flow response at small deformations is well demonstrated. In this section, we elaborate on training the flow response on experimentally measured data at large plastic deformations. One of the challenges involved in training with experimental data is the limitation of stress-strain data points as well as their frequency and form of occurrence through the deformation history. Access to the experimental data reported in the literature usually involves the extraction of data from stress-strain images using digitizing software. Typically data extraction is based on manual user input via the coordinates representing the image. Creating a pair of lists for the model output and measured data matching at the corresponding strain levels and necessary for a one-by-one comparison is challenging. Generating data with equal spacing is also not an option since adaptive time stepping is required for a computationally efficient framework with fewer time increments that represent the same curve with large deformations up to 10\% strain. 

\begin{figure}[H]
    \centering
    \includegraphics[scale=0.3]{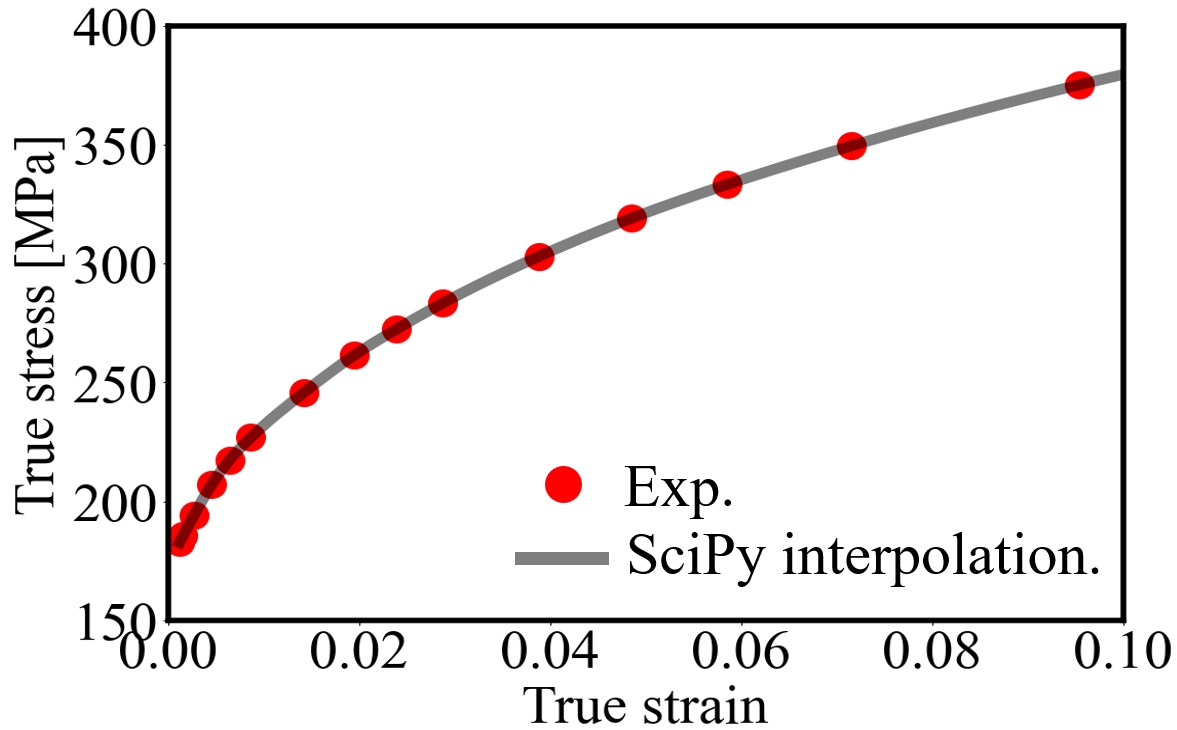}
    \caption{Interpolation of measured data in Ni using SciPy. }
    \label{fig:exp_copper}
    \end{figure}

\begin{figure}[H]
    \centering
    \includegraphics[scale=0.575]{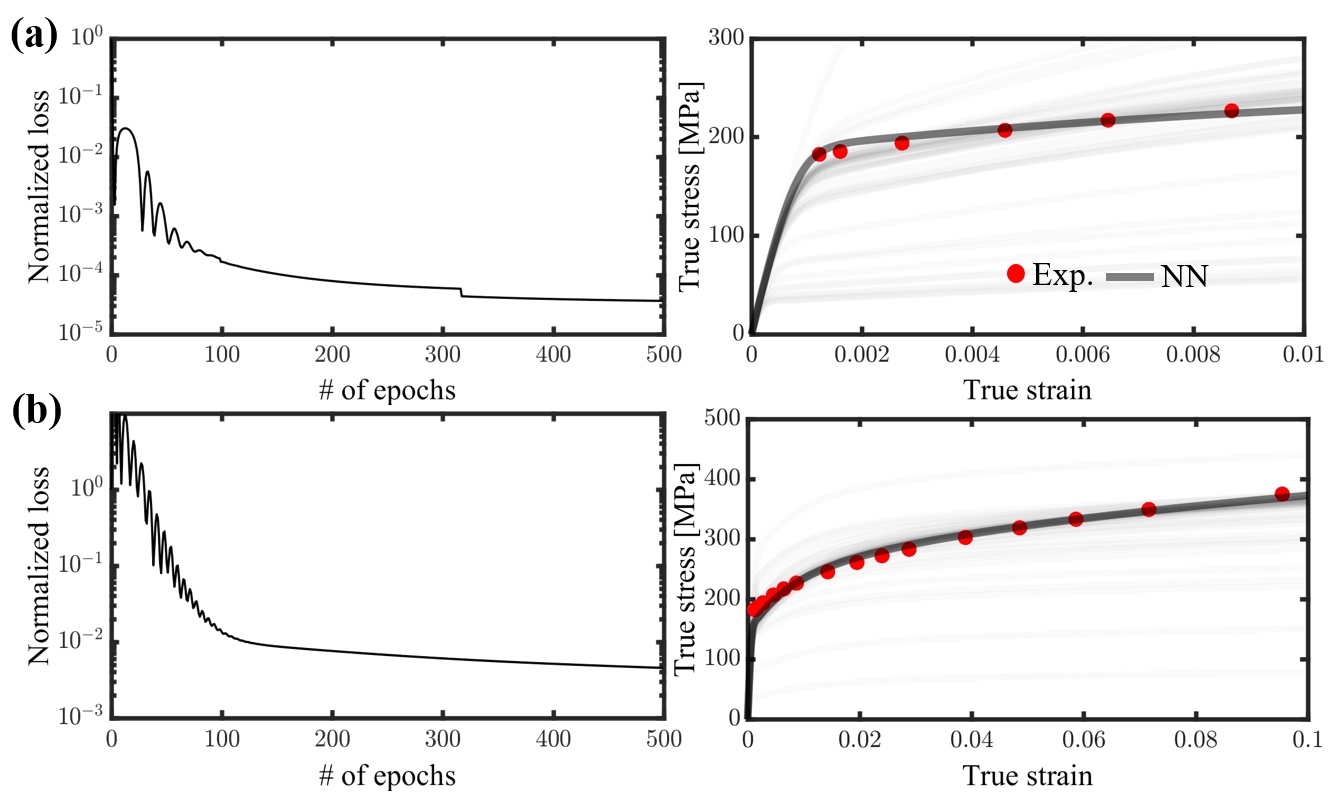}
    \caption{Model fit for measured flow response in Ni for (a) small deformation to a macroscopic strain of 1\% and (b) large deformation to a macroscopic strain of 10\%.} 
    \label{fig:exp_copper_fit}
    \end{figure}

Thus, to address this issue and to enable access to the experimental data points at arbitrary strain amplitudes with arbitrary and adaptive time stepping, a nonlinear interpolation preprocessing step via the Scipy \cite{virtanen2020scipy} library is used prior to training. Figure \ref{fig:exp_copper} shows an example of a continuous Scipy interpolation for experimentally obtained data in Ni with uneven spacing and arbitrary strain amplitudes. Once the Scipy interpolation is applied, the stress values corresponding to the desired strain levels are readily evaluated using the obtained interpolator. Figures \ref{fig:exp_copper_fit}(a) and \ref{fig:exp_copper_fit}(b) present the trained models using measured data for small and large deformations up to 1\% and 10\% strain in Ni. The model shows promising performance in capturing the elasto-viscoplastic transition in small strains as well as prediction of hardening curvature under large deformations. Note that since most of the experimental data existing in the literature ignore the elastic response, an additional step based on the current value of viscoplastic strain is imposed in algorithm \ref{alg:Train} to ignore the elastic response when computing the loss between the training output and measured response.      
\subsection{Grain size-aware flow response and Hall-Petch discovery}

\begin{figure}[H]
    \centering
    \includegraphics[scale=0.3]{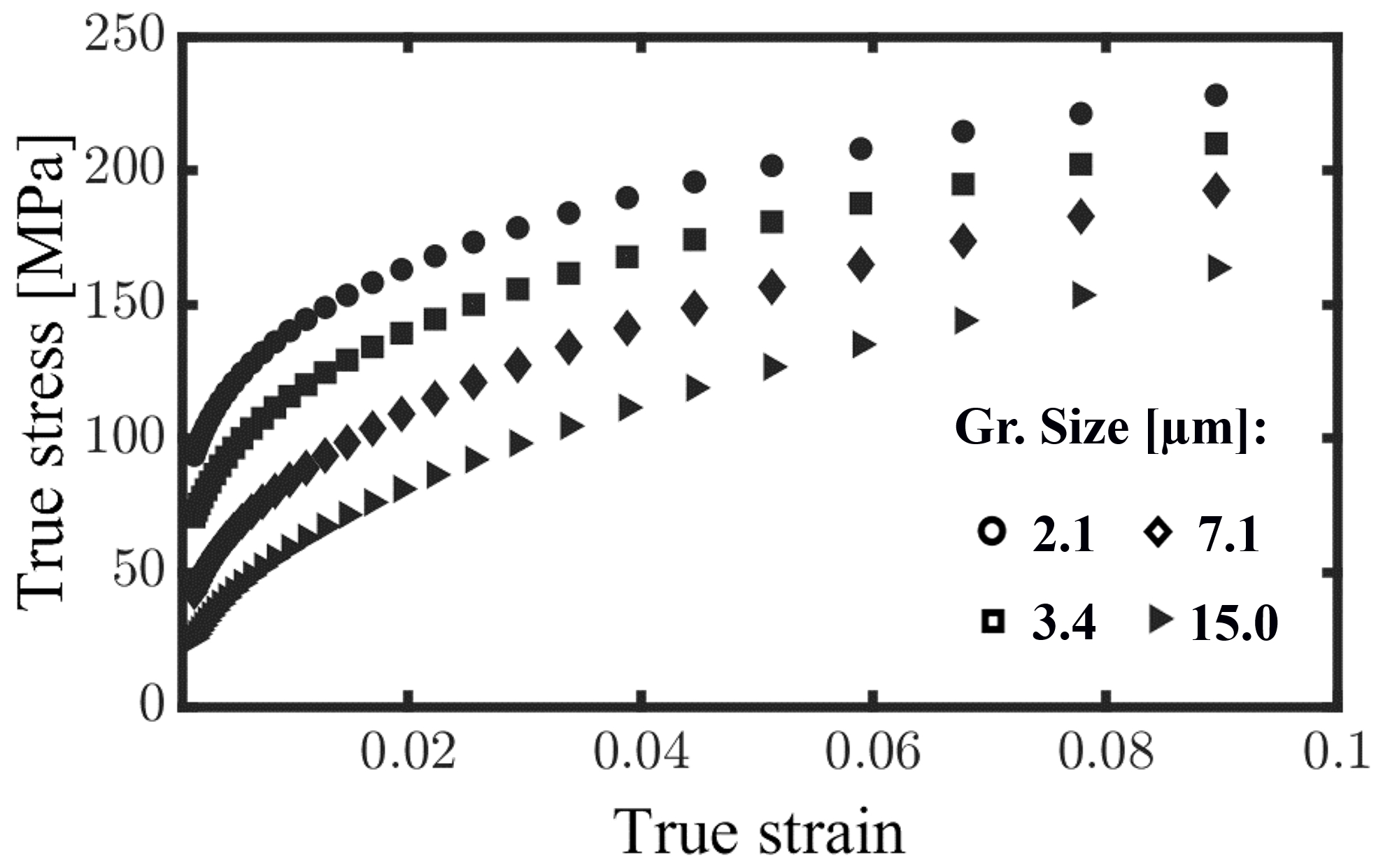}
    \caption{Variation in flow response of Cu under large deformation as a function of grain size.}
    \label{fig:HP_copper}
    \end{figure}

After validating the ability of the NN-EVP model in training experimental data to large strains with arbitrary data frequency, we take a step further to train the flow response at large deformations as a function of grain size and aim to discover the Hall-Petch relationship without considering a particular functional form. To this end, the measured flow response of Cu as a function of grain size and for average grain sizes of 2.1 $\mu m$, 3.4 $\mu m$, 7.1 $\mu m$ and 15 $\mu m$, as shown in Fig. \ref{fig:HP_copper}, is utilized for training the model including the grain size effects. Here, in addition to the dual potential network ${\mathcal{NN}_{\phi^{*}} (\sigma, R(r)) }$ and hardening network ${\mathcal{NN}_{R(r)}}$, the Hall-Petch neural network ${\mathcal{NN}_{HP(d_{grain})}}$ is also used in the model. In order to increase the training speed, adaptive time stepping was implemented with $\Delta t | _{t+\Delta t}=1.15 \Delta t | _t$ , resulting in 50 increments per stress-strain curve. Since the time stepping is adaptive, the training losses corresponding to each data point are scaled proportionally with the time increment to that data point to balance the loss evaluation. This is required due to the uneven data spacing resulting from irregular time steps where the density of training data is much smaller within the large deformation regime above 2\% strain.  

\begin{figure}[H]
    \centering
    \includegraphics[scale=0.75]{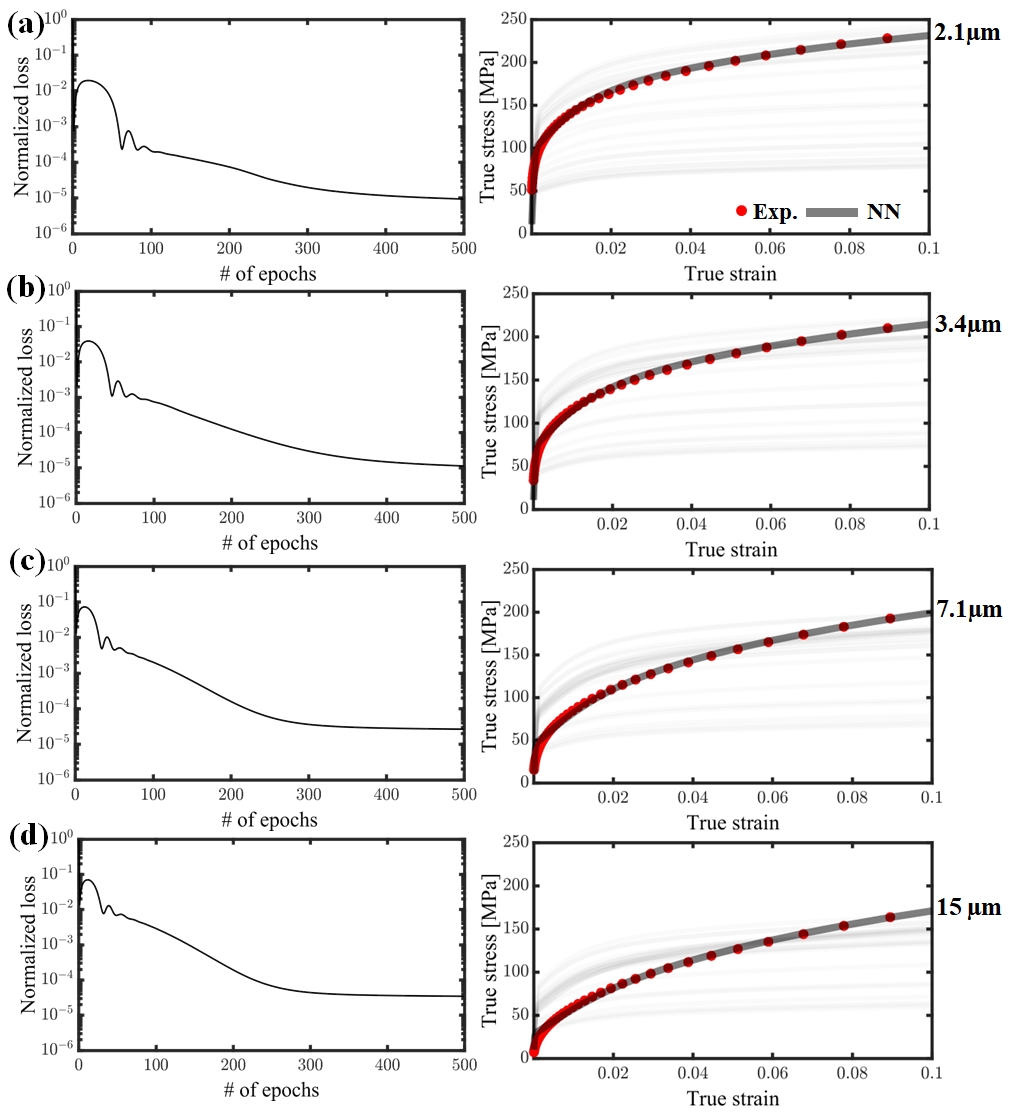}
    \caption{Single-fit training of flow response in Cu for grain sizes of (a) 2.1 $\mu m$, (b) 3.4 $\mu m$, (c) 7.1 $\mu m$ and (d) 15 $\mu m$.}
    \label{fig:HP_single_fits}
    \end{figure}

First, each curve is trained individually and independent of one another to test the ability of the model in training the flow response with varying stress levels. Results corresponding to single-curve fits are shown in Fig. \ref{fig:HP_single_fits}.  Since one curve is trained at a time, it is easier for the model to learn the flow response and thus the convergence is relatively fast. Next, a quaternary configuration (with all four stress-strain responses) including 200 data points is  utilized for training. Training on binary and ternary configurations with two and three curves consisting of 100 and 150 data points are also provided in the supplementary data shown in Figs. \ref{fig:HP_binary_fits} and \ref{fig:HP_ternary_fits}. Note that since multiple curves with contrasting stress levels and hardening curvatures are trained in parallel, it is more challenging for the model to converge to an optimal solution as evident in the behavior of loss evolution. Also, due to more data points the training process is computationally more demanding.       

\begin{figure}[H]
    \centering
    \includegraphics[scale=0.5]{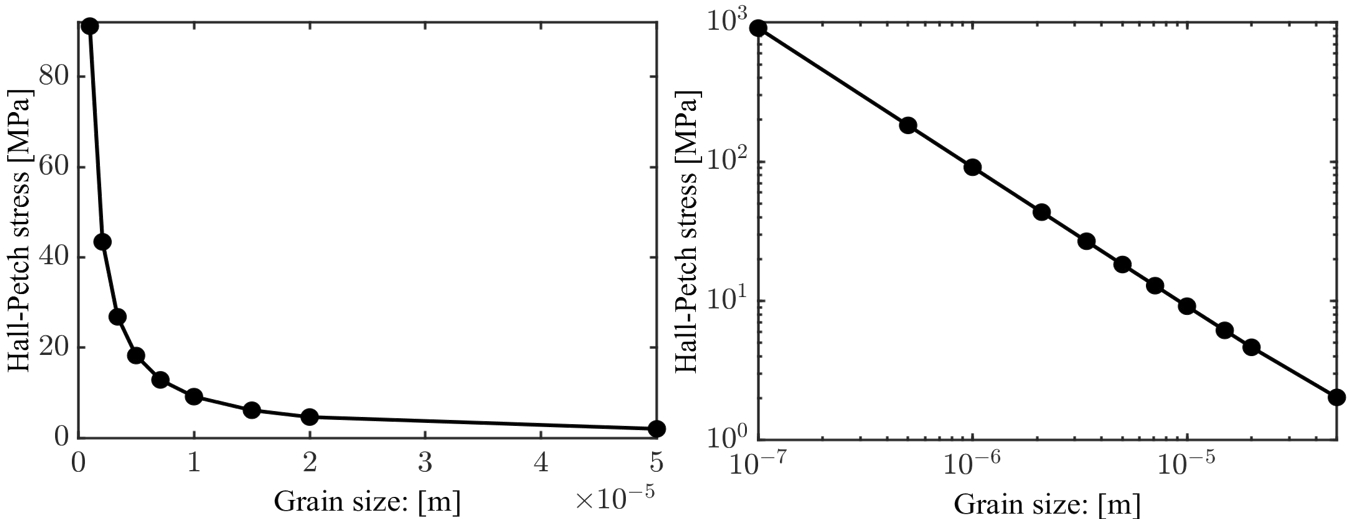}
    \caption{Discovery of Hall-Petch relationship in Copper using NN-EVP in non-logarithmic (on the left) and log-log (on the right) scales.}
    \label{fig:Hall_Petch_discovery}
    \end{figure}

Ultimately, we aim to discover the Hall-Petch relationship which describes the dependence of the initial yield on the grain size. To this end, after training the model for grain sizes of 2.1 $\mu m$, 3.4 $\mu m$, 7.1 $\mu m$ and 15 $\mu m$, the Hall-Petch neural network ${\mathcal{NN}_{HP(d_{grain})}}$ is recovered to extrapolate the initial flow response for grain sizes below  2.1 $\mu m$ and beyond 7.1 $\mu m$ with average grain sizes and their corresponding Hall-Petch stress values listed in Table \ref{tab:Hall_Petch_list}. The discovered Hall-Petch behavior is shown in Fig. \ref{fig:Hall_Petch_discovery}. These results indicate the ability of the model to capture the expected linear Hall-Petch relationship in log-log space, discovering it without any underlying assumptions on its functional form. Note that regardless of the particular material used herein, the proposed NN-EVP model should be able to predict the Hall-Petch relationship for a wide range of materials with arbitrary grain size and stress-strain response. Extrapolation of the Hall-Petch relationship is a complicated topic that we aim to study in the future.

        \begin{figure}[H]
    \centering
    \includegraphics[scale=0.475]{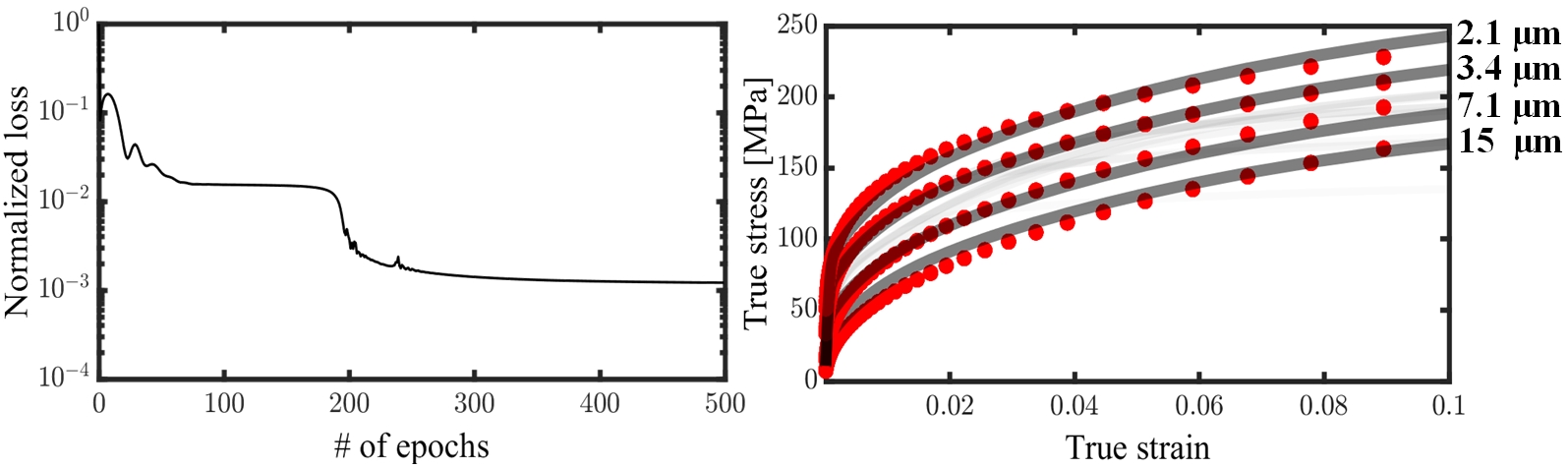}
    \caption{Multi-fit training of flow response in Cu for quaternary configuration of   2.1-3.4-7.1-15 $\mu m$.}   
    \label{fig:HP_quaternary_fits}
    \end{figure}

    \begin{table}
\begin{center}
\begin{tabular}{c c} 
 \hline
  Grain size $[\mu m]$  &  Hall-Petch stress $[MPa]$  \\ [0.5ex]  
 \hline  
 0.5 & 182.3671 \\ [0ex]
 \hline  
 1.0 & 91.1886 \\ [0ex] 
 \hline  
 2.1 & 43.4293 \\ [0ex] 
 \hline  
 3.4 & 26.8331 \\ [0ex] 
 \hline  
 5.0 & 18.2582 \\ [0ex] 
 \hline  
 7.1 & 12.8734 \\ [0ex] 
 \hline  
 10.0 & 9.1615 \\ [0ex] 
 \hline
 15.0 & 6.1435 \\ [0ex] 
 \hline
 20.0 & 4.645 \\ [0ex] 
 \hline  
 50.0 & 2.0286 \\ [0ex] 
 \hline  
 100 & 1.2712 \\ [0ex] 
  \hline  
 250 & 0.9902 \\ [0ex] 
  \hline  
 500 & 0.8907\\ [0ex] 
 \hline
\end{tabular}
\end{center}
\caption{Hall-Petch contributions in Cu via extrapolations of initial flow response as function of grain size.}\label{tab:Hall_Petch_list}
\end{table}

\section{Summary and conclusions }
The flow response in metals and metallic alloys differs drastically depending on the average grain size that forms the underlying microstructure. Modeling the viscoplastic flow response of metals is usually concomitant with the assumption of particular constitutive models in the form of a power law and phenomenological hardening formulations. These hardening laws often include parameters that involve manual calibrations with experimental data obtained from uniaxial tests. Machine learning models that can automate such calibrations generally rely on large amounts of data and often require a large number of iterations in order for them to be sufficiently accurate and predictive. For instance, the fitting process via genetic algorithms involve a large number of functional calls within four orders of magnitude to the black-box in order to obtain the flow response \cite{furukawa1997inelastic,jenab2016use}. To remove the need for these big datasets and large number of iterations in computationally expensive training, we propose a data-driven elasto-viscoplasticity framework based on neural networks called NN-EVP that leverages PyTorch's high-performance ML library. The proposed approach is tested and trained using both synthetic and experimentally measured uniaxial data in the context of limited data availability.  

The developed NN-EVP framework was adopted to train elasto-viscoplastic flow response of metals at both small and large deformations and as a function of grain size. First, rate sensitivity-dependent perfect viscoplasticity was discovered using the power law with no hardening effects. Next, synthetically generated deformation responses via a Johnson-Cook phenomenological hardening law were used to train and test the approach on its ability to extrapolate the flow response at strain amplitudes beyond the observed training data. Next, the framework was applied to train large deformations obtained from experimental tests with limited stress-strain data availability. Finally, simultaneous training of multiple grain size-dependent stress-strain curves enabled us to obtain a grain size-aware flow behavior and discover the Hall-Petch relationship pertaining to the grain size strengthening effects.      

The proposed NN-EVP model presented herein takes a further step in the prediction of flow responses in metals and improves the computational efficiency of structure-property-relationship simulations of metallic materials. The findings of the present work provide insights into the versatility and flexibility of data-driven constitutive modeling which can be adapted for a wide range of materials exhibiting complex large deformation behavior. This motivates future work in incorporating the current model into finite elements for efficient full-field modeling of large deformations under arbitrary loading and boundary conditions.

\section{Data availability }
Data and Python script supporting the findings of this study are available from the corresponding author upon request.
\section{Acknowledgments } 
JF, AE and NB gratefully acknowledge support by the Air Force Office of Scientific Research under award number FA9550-22-1-0075.  
 
 \clearpage
\appendix

\section{Supplementary data}\label{subsec:}

\begin{figure}[H]
    \centering
    \includegraphics[scale=0.675]{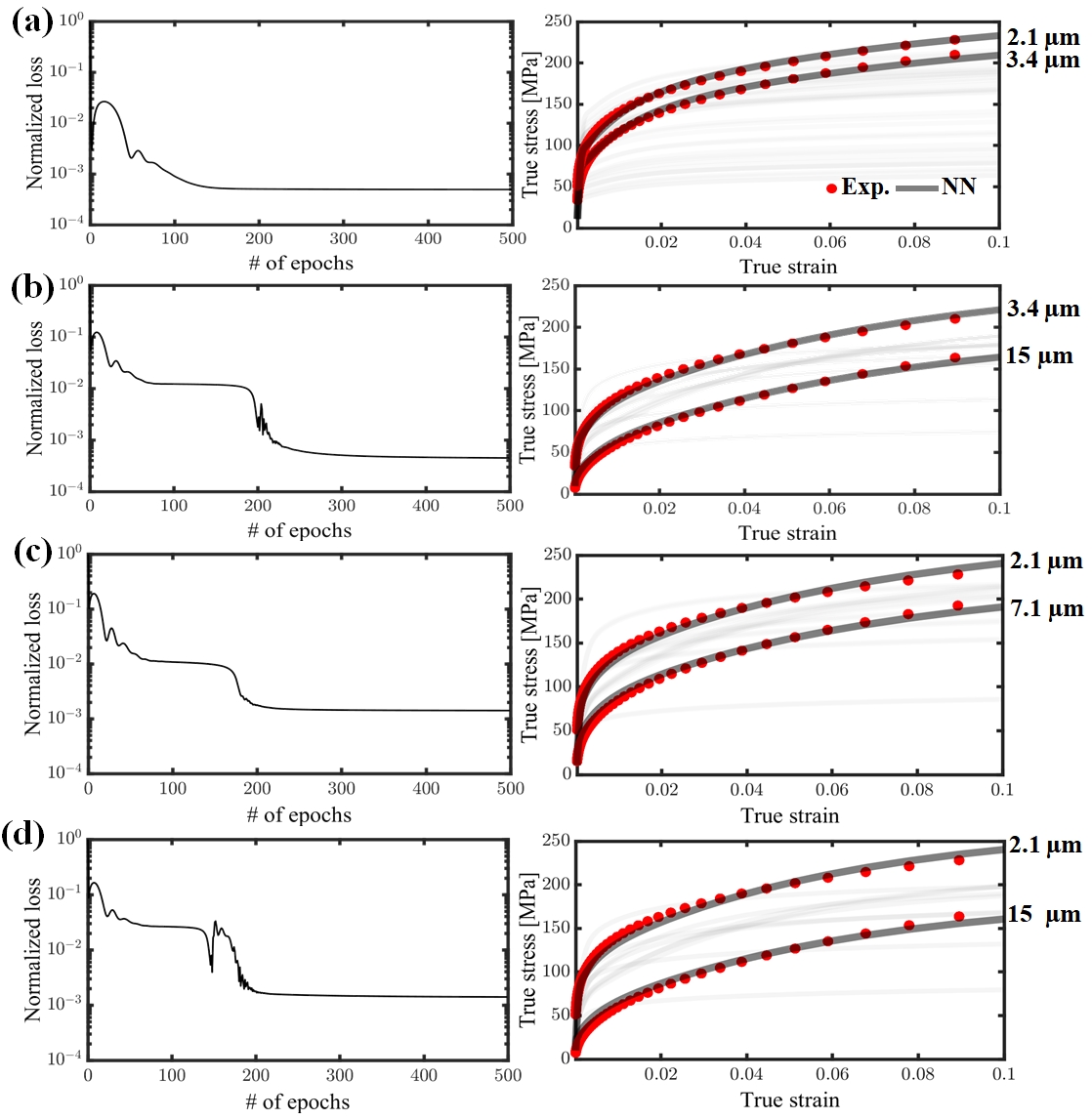}
    \caption{Multi-fit training of flow response in Cu for binary configurations of (a) 2.1-3.4 $\mu m$, (b) 3.4-15 $\mu m$, (c) 2.1-7.1 $\mu m$ and (d) 2.1-15 $\mu m$.}    
    \label{fig:HP_binary_fits}
    \end{figure} 
    \begin{figure}[H]
    \centering
    \includegraphics[scale=0.45]{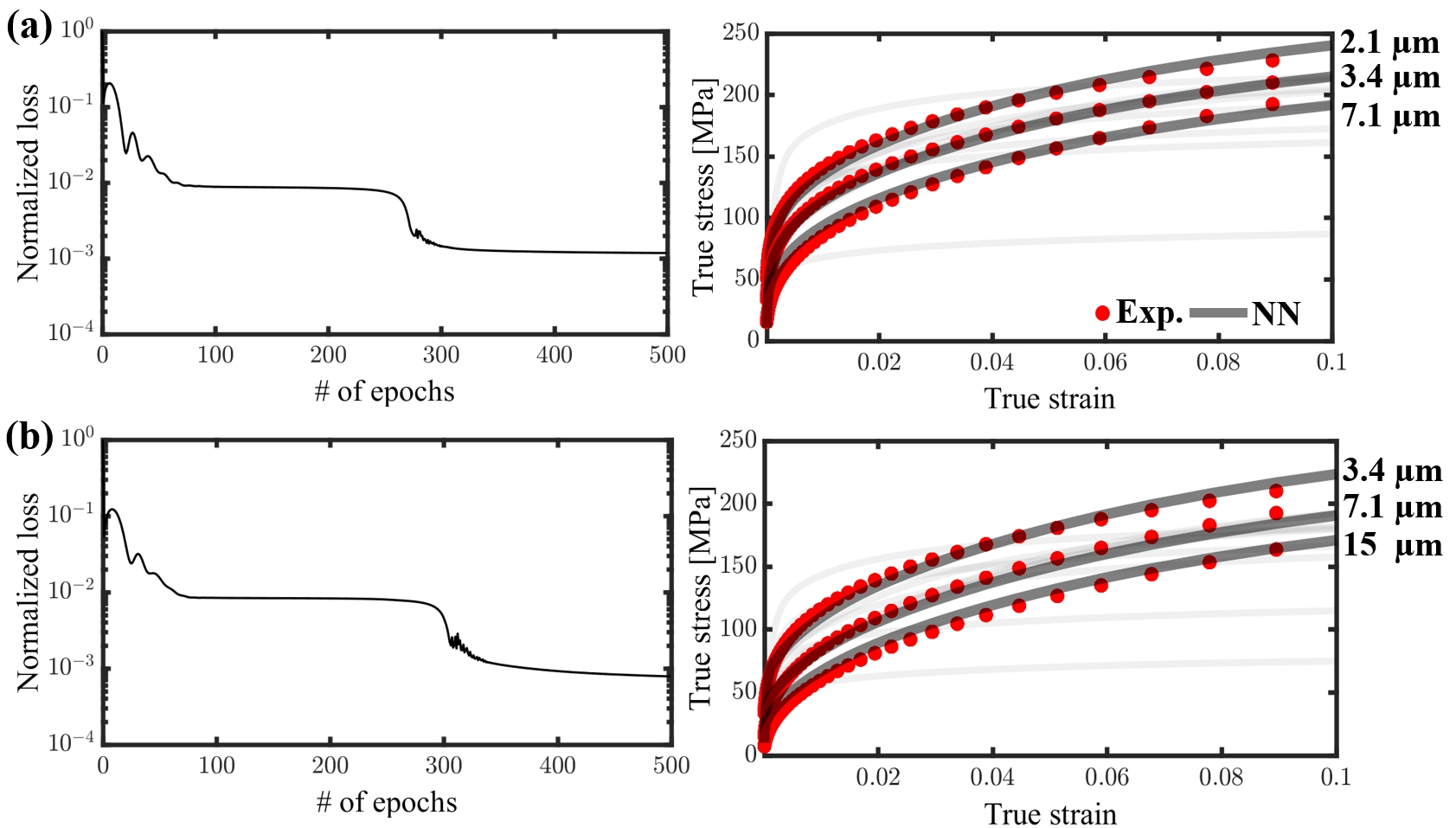}
    \caption{Multi-fit training of flow response in Cu for Ternary configurations of (a) 2.1-3.4-7.1 $\mu m$ and (b) 3.4-7.1-15 $\mu m$.}   
    \label{fig:HP_ternary_fits}
    \end{figure}

\break
\bibliography{References.bib}    
\end{document}